\documentclass[
reprint, 
 superscriptaddress,
]{revtex4-2}
\usepackage{amsmath}
\usepackage{graphicx}
\usepackage{dcolumn}
\usepackage{bm}
\usepackage[version=3]{mhchem} 
\usepackage[T1]{fontenc}       
\usepackage{amssymb}
\usepackage{txfonts}
\usepackage{xcolor}
\usepackage{xspace}
\usepackage{xr}
\usepackage{soul}
\usepackage{tikz}
\usepackage [mathscr] {eucal}

\usepackage[breaklinks=true,colorlinks,citecolor=blue,linkcolor=blue,urlcolor=blue]{hyperref}

\usepackage{tikz,xcolor,hyperref}

\definecolor{lime}{HTML}{A6CE39}
\DeclareRobustCommand{\orcidicon}{
	\begin{tikzpicture}
	\draw[lime, fill=lime] (0,0) 
	circle [radius=0.16] 
	node[white] {{\fontfamily{qag}\selectfont \tiny ID}};
	\draw[white, fill=white] (-0.0625,0.095) 
	circle [radius=0.007];
	\end{tikzpicture}
	\hspace{-2mm}
}
\foreach \x in {A, ..., Z}{\expandafter\xdef\csname orcid\x\endcsname{\noexpand\href{https://orcid.org/\csname orcidauthor\x\endcsname}
			{\noexpand\orcidicon}}
}

\begin{document}

\title{Phonon-Assisted Photoluminescence and Exciton Recombination in Monolayer Aluminum Nitride} 
\author{Pushpendra Yadav}
\email{pyadav@iitk.ac.in}
\affiliation{Department of Physics, Indian Institute of Technology Kanpur, Kanpur-208016, India}

\author{Amit Agarwal}
\email{amitag@iitk.ac.in}
\affiliation{Department of Physics, Indian Institute of Technology Kanpur, Kanpur-208016, India}

\author{Sitangshu Bhattacharya}
\email{sitangshu@iiita.ac.in}
\affiliation{Electronic Structure Theory Group, Department of Electronics and Communication Engineering, Indian Institute of Information Technology-Allahabad, Uttar Pradesh 211015, India}

\begin{abstract}
Efficient solid-state photon emitters with longer operating lifetimes in the ultraviolet (UV) wavelength range are crucial for optoelectronic devices. However, finding suitable material candidates has been a significant challenge. Here, we demonstrate that hexagonal aluminum nitride (AlN) monolayers exhibit strong photoluminescence emission within the UV range of $3.94-4.05$ eV. We show that these emissions in indirect bandgap AlN are facilitated by phonon modes with finite lattice momentum. These phonon modes promote efficient recombination of electrons and holes from the ${\Gamma}$ to ${\textbf K}$ point of the Brillouin zone. Our findings provide a foundation for developing advanced optoelectronic devices and efficient UV light sources based on hexagonal AlN monolayers. 
\end{abstract}
\maketitle
\section{Introduction}

Photoluminescence emission (PLE) experiments are crucial for investigating quantum phenomena and the behavior of optically excited carriers in semiconductors~\cite{Watanabe2004, Pozo-Zamudio_2015, Cassabois2016, Molas2017,Karmakar2021}. While PLE studies are well-established for bulk semiconductors, recent advancements have spurred a growing interest in exploring optical excitations in two-dimensional (2D) materials~\cite{Thygesen2017,Mounet2018,Santu2021, Yadav_2023_EHL-2D}. These studies provide deep insights into electronic band structures, excitons, and the effects of disorder and external fields, often revealing bound exciton complexes and indirect recombination processes~\cite{Wang2014PL, Brem2020}. To achieve a microscopic understanding of absorption and emission in quantum materials, a quantitatively accurate description is essential. Direct bandgap materials can be well-described by first-order perturbation theory. However, indirect bandgap semiconductors require a more complex approach that incorporates electron-phonon interactions (EPI) and second-order corrections~\cite{Cassabois2016}. A prominent example is the strong PLE observed in bulk hexagonal boron nitride (h-BN)~\cite{Watanabe2004}, where finite-momentum dark exciton states facilitate recombination through phonon modes~\cite{Cassabois2016}. Similar phonon-assisted PLE has been reported in h-BN encapsulated tungsten-based 2D dichalcogenides like WSe$_2$ and WS$_2$~\cite{Brem2020}.

EPI also play a key role in explaining the temperature-dependent optical properties of materials. Accurate descriptions require electronic energy states to be calculated using \textit{ab-initio} many-body perturbation theory (MBPT)~\cite{Marini2008}. While the GW approximation~\cite{HedinGW1,Hybertsen-GW2} effectively estimates electronic energies in excited states, it often fails to incorporate the effects of EPI and finite temperatures. Density functional perturbation theory addresses this by estimating the impact of EPI on electronic band structures and optical absorption spectra, especially in bulk semiconductors~\cite{Noffsinger2012,Mao2022}. For 2D insulators with high exciton binding energies, incorporating electron-hole interactions is crucial~\cite{SOC-effect-MoS2-PRL,Wang2018,Yadav_2023}. The temperature-dependent Bethe-Salpeter equation (BSE) effectively captures the impact of electron-hole correlations in the presence of phonons, successfully explaining the optical spectra of materials with varying exciton binding energies~\cite{Marini2008}. Among the 2D materials garnering recent interest, hexagonal aluminum nitride (h-AlN) stands out due to its high chemical stability, thermal conductivity, and mechanical strength~\cite{Zhuang13, Chandan2015, Dutta2016, Jiongyao2017, Michael2018}. These properties make h-AlN films highly promising for applications in solid-state optics, solar energy, and electronics~\cite{Jiang2017-GaN, AlN-exciton1}. Despite these advances, the emission processes in 2D h-AlN remain poorly understood.

Here, we study the impact of electron-phonon coupling on optical absorption and indirect emission in 2D h-AlN using \textit{ab-initio} MBPT. This approach has been successful in explaining phonon-assisted processes in conventional semiconductors such as Si, MoTe$_2$, and h-BN~\cite{Helmrich_2018, cannuccia2019theory, Noffsinger2012, Chen2020}. It allows us to analyze zero-momentum exciton lifetimes and demonstrate redshifted optical absorption spectra with decreased dipole oscillator strengths. We report shorter non-radiative (fs) and longer radiative (ps) exciton lifetimes, which significantly enhance quantum yield. 
Going beyond the optical dipole limit, our study reveals phonon-assisted indirect emissions in the UV range that persist at higher temperatures ($\ge$300 K). This results in an asymmetric spectrum around the optical gap, similar to h-BN, and is characterized by an indirect dark exciton at the \textbf{K} point of the center-of-mass momentum BZ~\cite{Ferreira_2023}. This dark exciton couples with phonon modes between \textbf{M} and \textbf{K}, playing a crucial role in excitonic thermalization and the emergence of phonon replicas in the emission spectrum within 3.94-4.05 eV. Our comprehensive analysis opens new avenues for optoelectronic applications in the ultraviolet regime.

\section{Temperature dependent Quasiparticle energies}

Absorption and PLE experiments are typically conducted at finite temperatures~\cite{Cassabois2016, Park2018,LAXMI2021148089}. To understand the renormalization of electronic states through phonon interactions at finite temperatures, we employ the dynamic Heine, Allen, and Cardona (HAC) theory~\cite{Allen1976, Allen1983}. Earlier studies have shown that lattice vibrations significantly impact the zero-point motion (ZPM), leading to notable renormalizations in exciton binding energies and linewidths~\cite{Ajay2023}. These effects have been benchmarked in a variety of semiconductors using fully \textit{ab-initio} methodologies. These include diamond-like materials~\cite{Lautenschlager1985, Lautenschlager1987, Zollner1992}, GaAs~\cite{Lautenschlager1987_}, bulk h-BN~\cite{Marini2008}, diamondoids~\cite{Gali2016}, cubic BN, LiF, and MgO~\cite{Antonius2015}, as well as atomically thin materials such as MoS$_2$~\cite{Chiara2020, Chan23}, WSe$_2$~\cite{HimaniWSe2}, h-BN~\cite{Mishra2019}, and NP~\cite{Kolos2021}.

\begin{figure}[t]
    \centering 
\includegraphics[width=1\columnwidth]{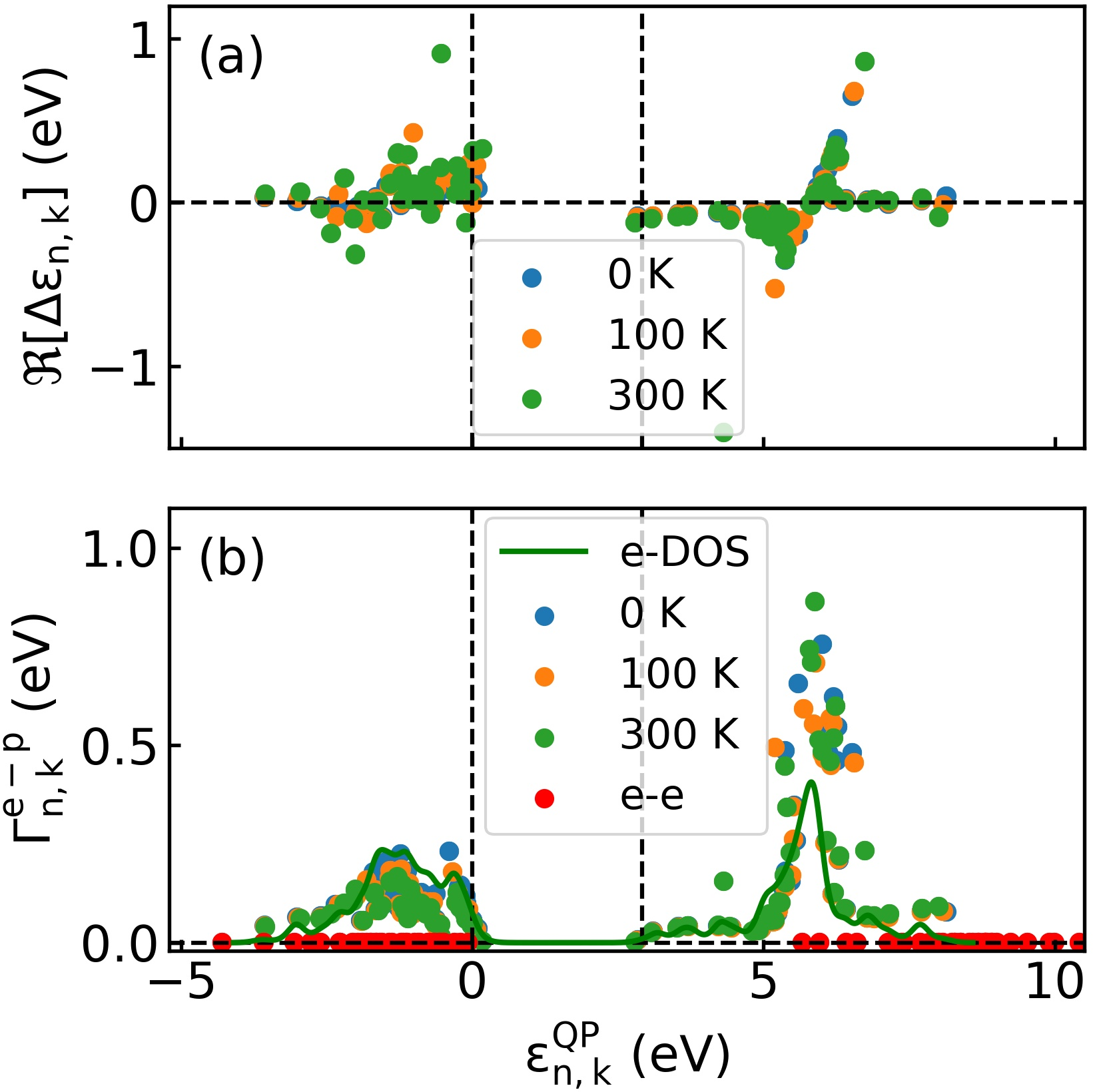}
    \caption{Renormalization of the electronic states by EPI.  (a) Effect of Fan+DW corrected energies at various temperatures exhibiting the strength of electron-phonon couplings. (b) Electronic linewidth modified by phonons at various temperatures. The corresponding phonon-assisted lifetime is $\hbar\left[2\Im\sum_{n\textbf{k}}^{\mathrm{Fan}}\left(\omega,T\right)\right]^{-1}$. The solid curve is the electronic DOS in the absence of lattice vibrations (scaled to fit), whereas the red dotted symbols are the GW corrected linewidths, including only electron-electron interaction effects and excluding phonon contributions.  
   \label{fig1}}
  \end{figure}

\begin{figure*}[t]
   \centering
 \includegraphics[width=0.8\linewidth]{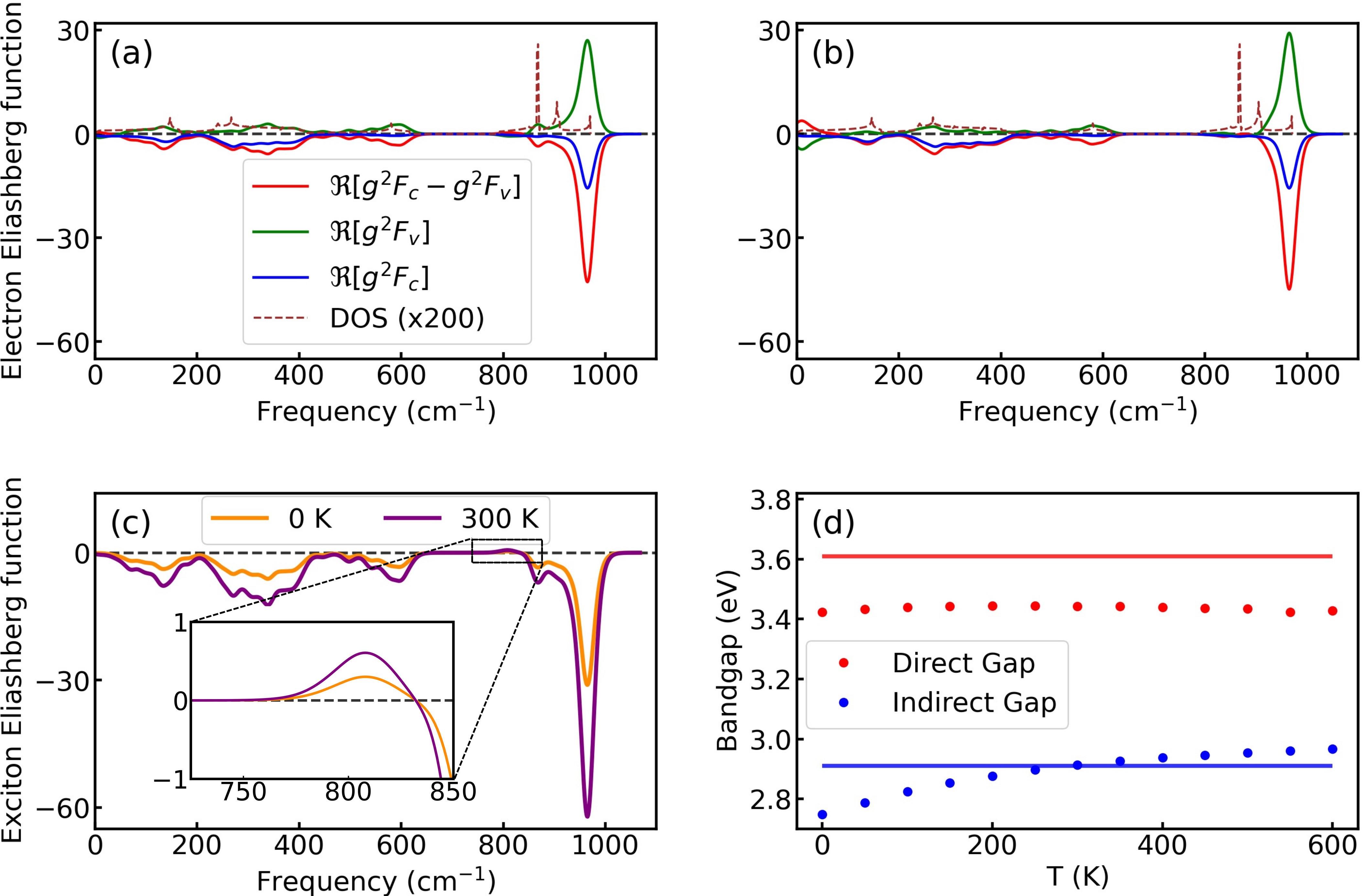}
   \caption{The electronic Eliashberg functions for the valence band maxima (green), conduction band minima (blue), and their difference (red) corresponding to the band edges for the (a) direct and (b) indirect band gap. (c) The exciton Eliashberg function at 0 K and 300 K and the inset to show the positive amplitude around 800 cm$^{-1}$ frequency. (d) The temperature-dependent electronic direct band gap (red dotted curve) and indirect band gap (blue dotted curve) with respect to the frozen atom condition direct band gap (red horizontal line) and indirect band gap (blue horizontal line). The direct band gap appears to be temperature independent in the 0-600 K range, while the indirect band gap shows a sub-linear increment with temperature. 
   \label{fig2}}
  \end{figure*}

2D h-AlN is a group-III nitride wide-bandgap semiconductor~\cite{ZhuangAlN}. We calculate its ground state electronic and thermodynamic stability using the density functional theory (DFT) and density functional perturbation theory (DFPT) based Quantum ESPRESSO code \cite{QE,GGA-PBE}. These are discussed in detail in the section SI and SV of the supplementary materials (SM)~\cite{Supplemental}. 
To estimate EPI and the renormalization of the electronic state $\left|n\textbf{k}\right\rangle$, we include both the Fan $\left(\textstyle \sum_{n\textbf{k}}^\mathrm{Fan}\left(\omega,T\right)\right)$~\cite{Fan1950} and Debye-Waller $\left(\textstyle \sum_{n\textbf{k}}^\mathrm{DW}\left(T\right)\right)$ ~\cite{Kim1986, Zollner1992} electron-phonon self-energies in the total electronic Hamiltonian. In Section SIV and SV of the SM~\cite{Supplemental}, we present the further details of the theoretical methodology and computation parameters. We present these significant polaronic corrections to electronic states in the figure~\ref{fig1}(a) which are otherwise absent in ground-state DFT analyses. The phonon-mediated electronic linewidths at different temperatures are proportional to the electronic density of states (e-DOS), and we present them in the figure~\ref{fig1}(b). We find that the phonon-renormalized states have significantly larger linewidths compared to the case considering only the dynamic correlation between electrons. Having obtained the self-energies, we evaluated the interacting electron's Green's function to construct the temperature dependent spectral function $\mathcal{A}_{n\textbf{k}}\left(\omega,T\right)$ for each state $\left|n\textbf{k}\right\rangle$~\cite{Lautenschlager1987_, Zollner1992, Marini2008}. The strength of the electron-phonon interaction can be understood from the shifts in the quasi-particle (QP) energies at various temperatures. We present a detailed discussion on $\mathcal{A}_{n\textbf{k}}\left(\omega,T\right)$, with calculations shown for the bottom conduction state (at $\Gamma$) and top valence (at $\Gamma$ and \textbf{K}) states in figure SIII, in the SM ~\cite{Supplemental}. We find that at 0 K, the valence states at $\Gamma$ and \textbf{K} blue-shift by about 111 meV and 85 meV, respectively. In contrast, the conduction state at $\Gamma$ red-shifts by about 79 meV from ground state (DFT) band-edge. This results in a net gap shrinkage of about 42 meV and 6 meV at the direct ($\Gamma$) and indirect ($\Gamma$-\textbf{K}) points, respectively. Figure \ref{fig2}(d) shows the variation of the electronic direct and indirect band gaps with temperatures. 
These shifts are also captured from the corresponding spectral functions shown in figure S5, and the renormalization of the electronic band gap (both direct and indirect) is summarized in table SI of the SM~\cite{Mahan2014, cannuccia2011, Supplemental}.

Next, we proceed to identify the phonon modes responsible for the renormalization of the band gap. For this, we use the polaronic quasi-particle (QP) energy ${\Delta\varepsilon_{n\textbf{k}}}$ to calculate the electronic Eliashberg function $g^{2}F_{n\textbf{k}}\left(\omega\right)$~\cite{Marini2008}. It  provides a microscopic description of the phonon modes involved in band gap renormalization, as shown in figure~\ref{fig2}(d). The function $g^{2}F_{n\textbf{k}}\left(\omega\right)$ is complex and it is summed over all phonon modes and branches $\textbf{q}\nu$. 
In figures~\ref{fig2}(a) and \ref{fig2}(b), we present the dependence of the real part of $g^{2}F_{n\textbf{k}}\left(\omega\right)$ with phonon frequencies at the valence band states at $\Gamma$ and \textbf{K}, and at the conduction band state at $\Gamma$. In conventional semiconductors, $\Re \left[g^{2}F_{v\textbf{k}}\left(\omega\right)\right]$ is positive at the top of the valence state, whereas $\Re \left[g^{2}F_{c\textbf{k}}\left(\omega\right)\right]$ is negative at the bottom of the conduction state. Thus, their difference $\Re \left[g^{2}F_{c\textbf{k}}\left(\omega\right) - g^{2}F_{v\textbf{k}}\left(\omega\right)\right]$ becomes negative, as shown in figure~\ref{fig2}(a). This cancellation is the key reason for band gap shrinkage with increasing temperature (see figure~\ref{fig2}(d)). The large negative area under this envelope [figure~\ref{fig2}(a)] justifies the decrease in the band gap with rising temperature. Additionally, this difference diminishes at both lower frequencies and the Debye frequency, signifying the validity of crystal translational invariance.

In the case of the indirect gap between $\Gamma$ and \textbf{K} point, the increment tendency could be due to the small positive amplitude of $\Re \left[g^{2}F_{c\textbf{k}}\left(\omega\right) - g^{2}F_{v\textbf{k}}\left(\omega\right)\right]$. 
Since the electronic Eliashberg function is proportional to the vibrational density of states (DOS), we can correlate the phonon modes responsible for these behaviors. Both electrons and holes are mainly coupled to frequencies around 965 cm$^{-1}$ of the optical branches. The small shoulder peak near 800 cm$^{-1}$ in figure~\ref{fig2}(a) is due to the in-plane longitudinal and transverse optical $\mathrm{E}$-type modes. In figure~\ref{fig2}(b), several frequencies where $\Re \left[g^{2}F_{c\textbf{k}}\left(\omega\right) - g^{2}F_{v\textbf{k}}\left(\omega\right)\right]$ becomes positive, such as near 900 cm$^{-1}$ and the lower acoustic modes around 10 cm$^{-1}$, are identified as major modes responsible for the band-gap increment at lower and higher temperatures.

Our findings highlight the impact of EPI on the electronic states in h-AlN and suggest that their inclusion is crucial for understanding optical excitations and photoluminescence. However, these are derived from a cascaded computational process with corrections of the order of a few milli-electron volts. As a consequence, the possibility of spurious computational inaccuracy cannot be completely ruled out. Carefully conducted experiments  
can help in establishing this behavior. 

\begin{figure}[t]
  \centering
  \includegraphics[width=0.9\linewidth]{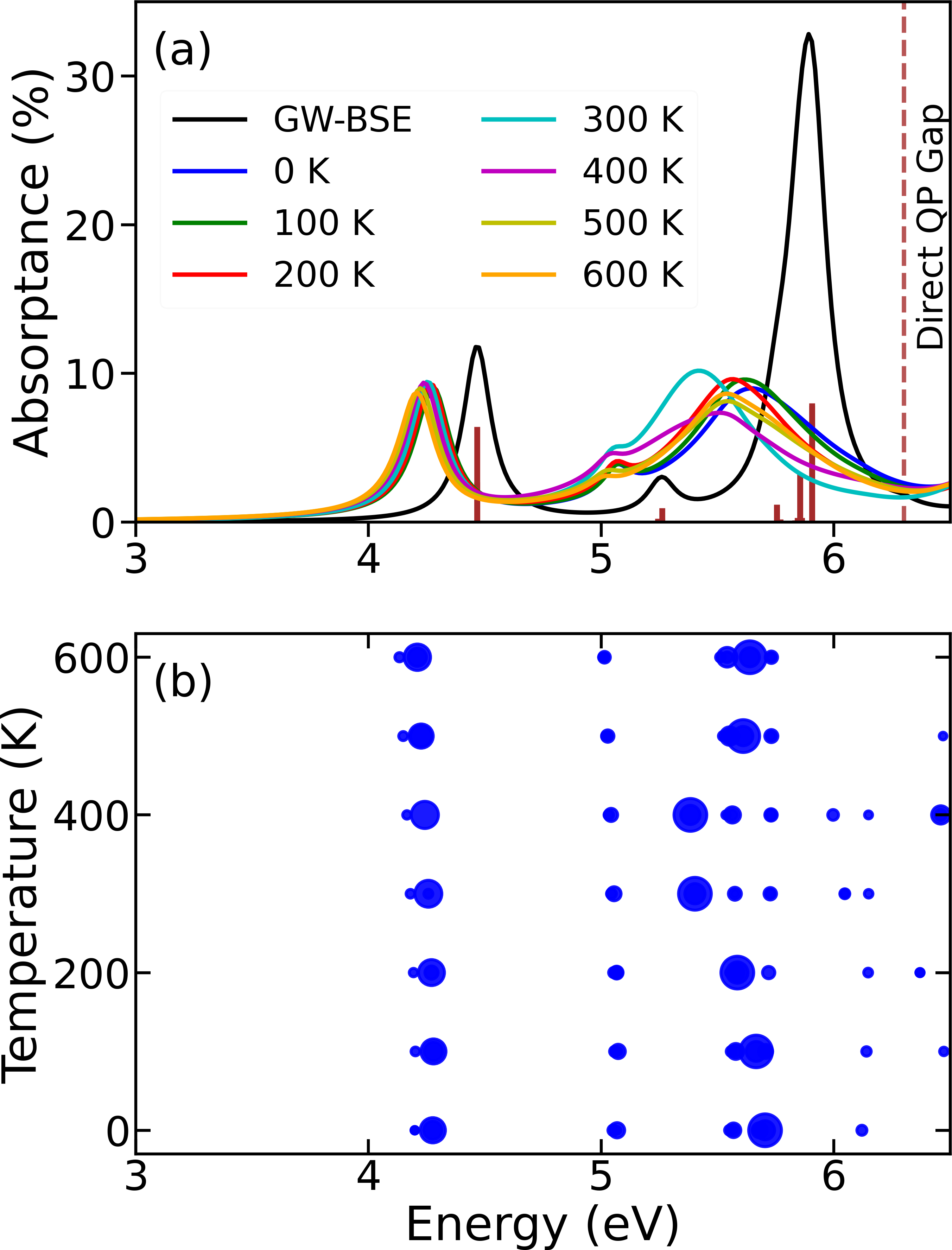}
  \caption{The impact of lattice vibrations on the absorption, excitonic structure, and excitonic non-radiative linewidths. (a) The absorption in the frozen atomic configurations and at different temperatures. Due to the lattice vibrations, the EPI modify the excitonic resonances largely at 0 K, whereas the changes at finite temperature compared to 0 K are small. (b) The exciton dipole oscillator strength. The dipole oscillator strength for the first exciton (4.47 $eV$ ) increases with temperature, but the second bright exciton (5.26 $eV$) becomes a dark exciton after 200 $K$. Interestingly, the third bright exciton (at 5.90 $eV$) shows a redshift-blueshift crossover across the different temperature ranges. We observe that the lattice vibrations significantly influence excitonic properties, altering absorption, and dipole oscillator strength with notable temperature-dependent transitions and shifts in excitonic behavior.}  
  \label{fig3}
  \end{figure}
\section{Temperature dependent optical absorption}
We now study the impact of lattice vibrations and on the optical absorption of 2D h-AlN. We start with the standard two-body Bethe-Salpeter (BS) Hamiltonian~\cite{BSE-1-Strinati, Rohlfing2000,yambo2019}, 
\begin{equation} \label{bse}
{
\left(\varepsilon_{c\textbf{k}}-\varepsilon_{v\textbf{k}}\right)A_{vc\textbf{k}}^{s}+
\sum_{v^{\prime}c^{\prime}\textbf{k}^{\prime}}\left\langle vc\textbf{k}\left|\mathrm{K}_{vc\textbf{k},v^{\prime}c^{\prime}\textbf{k}^{\prime}}\right|v^{\prime}c^{\prime}\textbf{k}^{\prime}\right\rangle A_{v^{\prime}c^{\prime}\textbf{k}^{\prime}}^{s}=\mathcal{E}_{X}^{s}A_{vc\textbf{k}}^{s}
}~.
\end{equation}
Here, ${\varepsilon_{c,v\textbf{k}}}$ are the QP energies, ${A_{vc\textbf{k}}^{s}}$ is the excitonic amplitude in state $s$ with electron and hole states $\left|c\textbf{k}\right\rangle$ and $\left|v\textbf{k}\right\rangle$, respectively. ${A_{vc\textbf{k}}^{s}}$ and exciton energies ${\mathcal{E}_{X}^{s}}$ are obtained by diagonalizing equation (\ref{bse}) with the electron-hole interaction kernel $\mathrm{K}$. We calculate the QP energies by computing the dynamical electron-electron self-energies~\cite{Aryasetiawan1998} using the GW approximation. The resulting electronic band structures using DFT and GW  are presented in figure~\ref{fig6}(c). We find that monolayer h-AlN has an indirect gap of about 5.73 eV located between the $\Gamma$ and \textbf{K} points in the BZ, while the direct gap, located at $\Gamma$, is 6.30 eV, within the GW approximation.

 The optical absorbance spectrum calculated both at the GW-BSE level and in the absence of interaction kernel, \textit{i.e.} the independent particle (IP) approximation, are shown in figure~\ref{fig3}(a). The spectrum in the presence of the kernel shows major excitonic peaks in the 4.0-6.0 eV range. This is in contrast to the step-like absorbance with peak at the edge of the QP gap within the IP approximation. The first exciton peak, at 4.47 eV, corresponds to the fundamental optical band gap. The exciton binding energy for this lowest optically bright energy exciton (say B$_1$ exciton) is 1.83 eV, indicating a strong electron-hole correlation. Two more prominent peaks appear at 5.86 eV (shoulder peak) and 5.91 eV (B$_2$ and B$_3$ excitons, respectively) below the QP direct band gap. These three peaks consist of doubly degenerate bright pairs with large exciton oscillator strength. We present the excitonic wavefunction for these three prominent excitons in figure (S2) of the SM~\cite{Supplemental}. 

Monolayer h-AlN possesses a C$_{3v}$(3m) point group symmetry with three irreducible representations: A$_1$, A$_2$, and $E$. The former two are one-dimensional with even and odd $\sigma_{v}$ reflection symmetries, respectively, while the last irreducible representation $E$ is two-dimensional. We find that, all three doubly degenerate bound excitons have an $E$-type symmetry. Figures S2 (d)-(f) in SM demonstrate this symmetry, with negligible electronic charge densities being present at the hole site. The trigonal character of the excitons observed, consistent with an overall C$_{3v}$(3m) symmetry. The lowest bound exciton is tightly bound and spread out only to the nearest sites, suggesting a Frenkel exciton character. In contrast, the next two excitons exhibit characteristics similar to Mott-Wannier excitons with relatively larger exciton radius. 
We find that all excitons proliferate along the aperiodic direction, with increasing volume increasing at higher energies. This behavior becomes non-trivial for excitonic interactions between bilayers or hetero-structures, leading to additional Davydov splittings in the absorption spectrum~\cite{Paleari2018, Henrique2017}.

To understand the impact of finite temperature on the optical excitations, we incorporate electron-phonon interaction. Unlike the frozen atom condition, where atoms are assumed to be fixed at lattice points making the BS Hamiltonian to be Hermitian, the inclusion of EPI, transforms this Hamiltonian into a non-Hermitian matrix~\cite{Marini2008}. Consequently, the excitonic energy eigenvalues become complex quantities. The correction to the real part of the exciton eigenvalues can be expressed as, 
\begin{equation}
\begin{split}
\Re\Delta \mathcal{E}^{\mathrm{s}}(T) = & \left[\left\langle \mathrm{s}(T)\left|H^{\mathrm{FA}}\right|\mathrm{s}(T)\right\rangle - \left\langle \mathrm{s}\left|H^{\mathrm{FA}}\right|\mathrm{s}\right\rangle\right] \\
& + \int d\omega \Re\left[g^{2}F_{\mathrm{s}}(\omega, T)\right] \left(n_{B}(\omega, T) + \frac{1}{2}\right)~.
\end{split}
\end{equation}
Here, $n_B$ denotes the Bose occupation factor. The quantity $g^{2}F_{\mathrm{s}}\left(\omega,T\right)$, known as the exciton-phonon coupling function (or excitonic Eliashberg function), is analogous to the electronic Eliashberg function and is proportional to the temperature-dependent excitonic amplitudes. 
The broadening of the exciton energy is induced by the coherent interaction between the exciton and the phonon, as well as the individual interactions of the electron-phonon and hole-phonon. 

The temperature dependent optical absorption spectrum, defined as $A=1-\textrm{exp}\left(-\frac{\Im\varepsilon\left(\omega, T\right)\mathcal{E_{\mathrm{s}}}d}{\hbar c}\right)$, is obtained by utilizing the temperature-dependent BSE~\cite{Marini2008,Molina2016}. Here, $\Im\varepsilon\left(\omega, T\right)=\frac{4\pi\alpha_{2D}\left(\omega, T\right)}{d}$ is the imaginary part of the temperature-dependent dielectric function, $d$ and $c$ are the 2D AlN thickness and speed of light respectively. We present the calculated optical absorption in figure~\ref{fig3}(a). We find that the spectra at different temperatures are significantly red-shifted compared to the frozen atom GW-BSE result. 
The fundamental exciton peak at 0 K is red-shifted by approximately 200 meV, indicating the effect of ZPM capturing the intrinsic spatial uncertainty of atoms respecting Heisenberg's uncertainty principle. 
The full-width at half maximum (FWHM) represents the phonon-assisted broadening in the optical spectra. The red-shifting of the exciton peaks can be understood from the sign of $g^{2}F_{s}\left(\omega,T\right)$ where a net cancellation between the coherent and incoherent terms can result in the lowering or increasing of real exciton energies~\cite{Molina2016,Marini2008}. For instance, if $g^{2}F_{c}\left(\omega, T\right) > g^{2}F_{v}\left(\omega, T\right)$, the area under $g^{2}F_{s}\left(\omega,T\right)$ is positive (see figure ~\ref{fig2} (c) and the inset), leading to a blue-shift in the spectrum. Conversely, if $g^{2}F_{c}\left(\omega\right) < g^{2}F_{v}\left(\omega\right)$, the area becomes negative, resulting in a red-shift in the spectrum. This behavior is demonstrated in figure~\ref{fig2}(c), where we observe that the incoherent interaction with the phonons causes $g^{2}F_{s}\left(\omega,T\right)$ to be mostly negative at all frequencies. The most prominent peak is located near 900 cm$^{-1}$, and it corresponds to the optical phonon branches at $\Gamma$. As temperature increases, the phonon number density rises, with the most intense interaction coming from branches near 900 cm$^{-1}$. These are the modes where the torsional motion of atoms is ubiquitous, thereby stretching and compressing the 2D sheet, relaxing and reinforcing the excitonic interactions with the lattice. 

 The intense exciton-phonon interaction can be further understood by analyzing the dipole oscillator strength. Figure~\ref{fig3}(b) shows the exciton energies weighted by their oscillator strengths as a function of temperature. We consider only those excitons whose strengths are more than 10$\%$ of the maximum, designating them as bright. The effect of temperature is more evident in the states of these excitons. Corresponding to the temperature-dependent absorption, we observe a slow red-shifting of the fundamental bound exciton near 4.25 eV, indicating that it remains bright throughout the entire temperature range (as shown by the size of the circle). Similarly, the other bound exciton near 5.06 eV experiences a slower red-shifting with comparatively less strength and, therefore, appears darker. A relatively faster red-shifting of the exciton energies with temperature is observed for the group of bound excitons in the energy range of 5.25-5.75 eV. Interestingly, a few excitons near 5.06 and 5.25 eV share their oscillator strengths when in close proximity. Their strengths increase when two excitons are close by, whereas they both lose strength and become dark when they separate. This exchange of optical strength could be due to coherent interactions, similar to those observed in Si and h-BN~\cite{Marini2008}.

\section{Non-radiative and radiative linewidths of excitons}
\begin{figure}[t]
  \centering
  \includegraphics[width=1\linewidth]{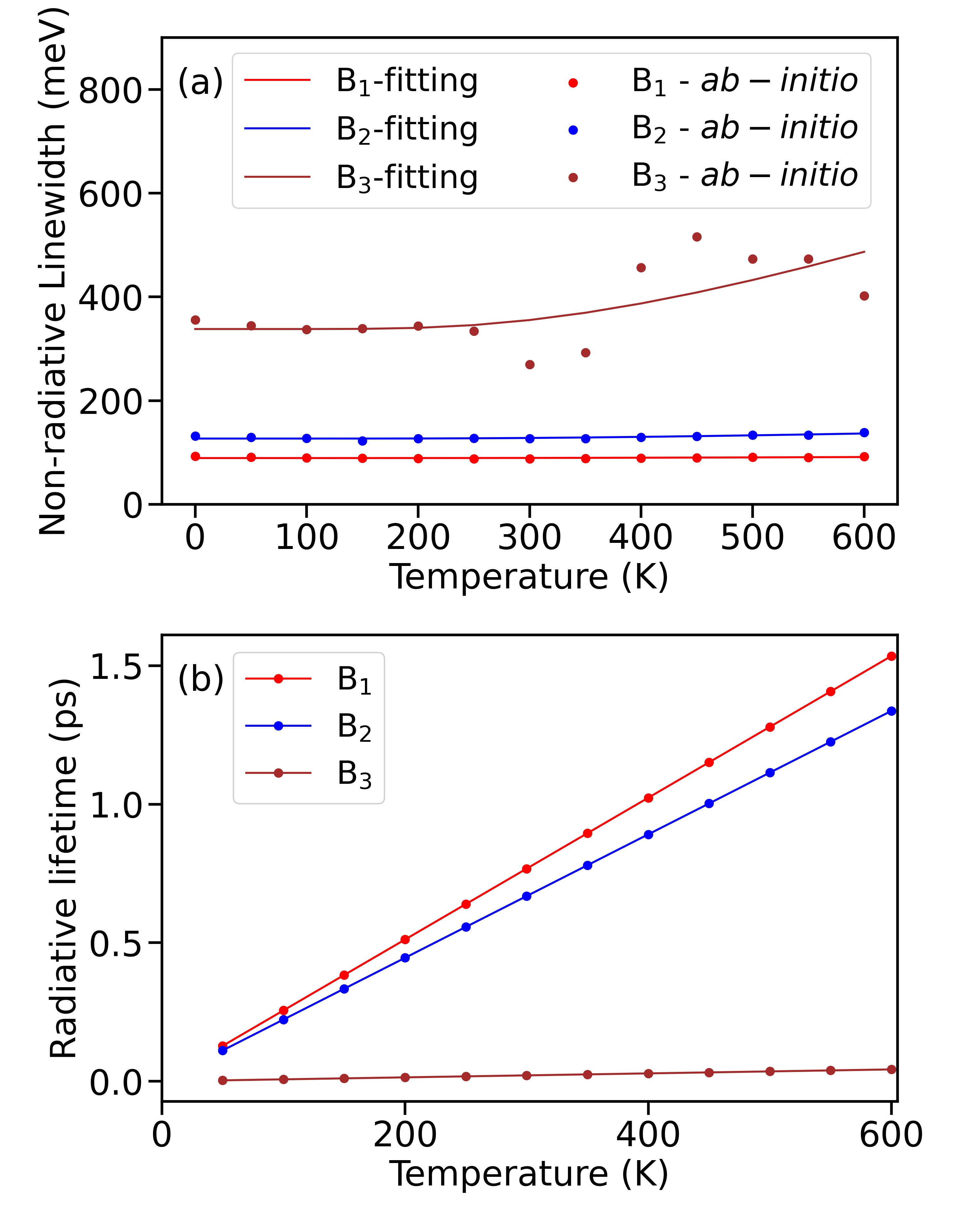}
  \caption{ (a) The excitonic linewidth for non-radiative processes, calculated from the \textit{ab-inito} method and fitted with the empirical relation (see equation ~\ref{eq_nr2}), and (b) lifetime for the radiative recombination of the three prominent bright excitons (B$_1$, B$_2$, and B$_3$ excitons) as a function of temperature. Excitonic linewidths for the non-radiative  process increases with temperature, whereas, for the radiative process, the linewidth (lifetime) decreases (increases) with temperature.} 
  \label{fig4}
 \end{figure}

To understand finite excitonic lifetimes exciton such as nonradiative dynamics~\cite{Purz2022}, decoherence times~\cite{Moody2015,dey2016optical}, we utilize the imaginary part of the exciton eigenvalues which are called exciton linewidths~\cite{Selig2016,cadiz2017excitonic}. The excitonic nonradiative recombination rate can be presented as ($\gamma_{NR}$)~\cite{Marini2008},
\begin{equation}
\begin{split}
\gamma_{NR}\left(T\right) = &\left(2\tau_{NR}^{\mathrm{s}}\right)^{-1}\\
= &\int d\omega\Im\left[g^{2}F_{\mathrm{s}}\left(\omega,T\right)\right]\left[n_{B}\left(\omega,T\right)+\frac{1}{2}\right]~   \label{eq_nr}
\end{split}
\end{equation}
Here, $\tau_{NR}^s$ denotes the nonradiative lifetime of the $s^{th}$ exciton. We present the non-radiative excitonic linewidths of the three prominent bright excitons (B$_1$, B$_2$, and B$_3$) in figure~\ref{fig4}(a). At 0 and 600 K, the linewidths lie between 60 meV and  500 meV. As the optical branches of phonons dominate the interactions, we estimate the nonradiative recombination rates using the empirical equation \cite{Selig2016},
\begin{equation}
\gamma_{NR}\left(T\right)=\gamma_{0}+\gamma_{\textrm{op}}\left[\textrm{exp}\left(\frac{\Lambda}{k_{B}T}\right)-1\right]^{-1}~.
\label{eq_nr2}
\end{equation}
Here, $\gamma_{0}$ denotes the zero Kelvin residual linewidth, $\gamma_{\textrm{OP}}$ represents the interaction strength between excitons and optical phonons and $\Lambda$ denotes the relevant phonon frequency. Our calculation for monolayer h-AlN indicate a $\gamma_{0}$=89.21 meV, at $\gamma_{op}$=13.09 meV.  We find that in AlN, there is limited impact of acoustic phonons, and these linewidths increase significantly with temperature after about 350 K. This makes $\gamma_{NR}$(300 K)$\sim\gamma_{0}$, which indicates a strong exciton-optical phonon coupling in AlN. This 
is in stark contrast to CVD grown capless WSe$_2$ on sapphire substrate, where the prominent interaction comes from acoustic branches~\cite{Moody2015}. However, in case of an h-BN encapsulated MoS$_2$~\cite{cadiz2017excitonic} the interaction at at higher temperatures is mainly dominated by optical phonons.

Within the correlated electron-hole picture, the exciton energies and dipole oscillator strength could further be utilize to understand the intrinsic radiative recombination lifetime to compare it with the phonon-assisted non-radiative lifetime using~\cite{Chen2018}. The radiative recombination rate is given by,  
\begin{equation}
\ensuremath{\gamma_{R}^{\mathrm{s}}=\left[\tau_{R}^{\mathrm{s}}\right]^{-1}=\frac{e^{2}\mathcal{E}_{\mathrm{s}}}{\epsilon_0\hbar^{2}c}\frac{\mu_{\mathrm{s}}^{2}}{A_{uc}}}~.   
\label{eq:radiative}
\end{equation}
Here, $A_{uc}$ is the primitive cell area, $\epsilon_0$ is the free space permitivity, $\hbar$ reduced Planck's constant, and $e$ magnitude of electronic charge. In equation~\eqref{eq:radiative}, $\mu_{\mathrm{s}}^{2} = |\Sigma_{cv\textbf{k}}A^{s}_{cv\textbf{k}} \langle \phi_{c\textbf{k}}|\textbf{r} |\phi_{v\textbf{k}}\rangle |^2/N_{k}$ denotes the $s$-exciton intensity as the linear combination of the square of the transition matrix elements between electron-hole pairs with the corresponding amplitudes $A^{s}_{cv\textbf{k}}$ divided by the sampled momenta $N_k$~\cite{Chen2018,Chen2019,Palummo2015,Nilesh2024}. 

\begin{table}[!t]
\caption{Excitonic radiative and non-radiative lifetimes in 2D h-AlN at 0 K and 300 K.} 
\label{table:LW}
\centering
\begin{ruledtabular}
\begin{tabular}{cccccc}
Excitons & $\tau_{NR}$ & $\tau_{NR}$ & $\tau_{R}$ & $\left\langle \tau_{R}\right\rangle$ & $\left\langle \tau_{R}^{eff}\right\rangle$ \\
 (Energy in eV)& (0 K) & (300 K) & (0 K) & (300 K) & (300 K) \\
 & (in fs) & (in fs) & (in fs) & (in ps) & (in ps) \\
\hline
B$_1$ (4.47)  & 2.33 & 2.33 & 3.99 & 0.77 & 0.76 \\
B$_2$ (5.86)  & 1.64 & 1.63 & 5.96 & 0.67 & 0.66 \\
B$_3$ (5.91)  & 0.61 & 0.58 & 2.47 & 0.02 & 0.02 \\
\end{tabular}
\end{ruledtabular}
\end{table}
At low temperatures, a thermally averaged radiative lifetime can be formulated as 
\begin{equation}
\ensuremath{\ensuremath{\ensuremath{\left\langle \tau_{R}^{\mathrm{s}}\right\rangle =\tau_{R}^{\mathrm{s}}\frac{3}{4}k_{B}T\left(\frac{2m_{\mathrm{s}}c^{2}}{\mathcal{E}^{2}_{\mathrm{s}}}\right)}}}~, 
\label{eq_rad}
\end{equation}
 under the parabolic exciton dispersion assumption which is linearly proportional to temperature~\cite{Chen2019}. Here, $k_B$ is the Boltzmann constant, $m_s$ is the effective exciton mass. 
 Moving to higher temperatures, an effective lifetime can be defined as, $\left\langle \tau_{R}^{eff}\right\rangle ^{-1}=\frac{\sum_{\mathrm{s}}\left\langle \tau_{R}^{\mathrm{s}}\right\rangle ^{-1}e^{-\mathcal{E}^{\mathrm{s}}/k_{B}T}}{\sum_{\mathrm{s}}e^{-\mathcal{E}^{\mathrm{s}}/k_{B}T}}$~\cite{Palummo2015}, which is reasonably close to experimental observations and is weighted over the averaged lifetime of degenerate excitons. The radiative lifetime of excitons are shown in figure~\ref{fig4}(b). We find that at 0 K, the intrinsic lifetime of the three prominent bound excitons ranges from about 3.99 fs to 5.96 fs. However, this dramatically increases to the order of picoseconds at higher temperatures, primarily due to the large oscillator strength, exciton energies, etc. These lifetimes are comparable to those observed in 2D TMDCs~\cite{Palummo2015}. When comparing to the $\tau_{NR}$, we observe that it is the phonon-assisted non-radiative processes which are significantly faster (in fs), even at 300 K, leading to a substantial ratio $\frac{\left\langle \tau_{R}\right\rangle}{\tau_{NR}}\sim10^{3}$. Such ultrafast non-radiative lifetimes are also observed experimentally in various 2D TMDCs~\cite{Selig2016} and therefore indicate a significantly enhanced quantum yield $\left(\frac{\left\langle \tau_{R}\right\rangle }{\left\langle \tau_{R}\right\rangle +\tau_{NR}}\right)\rightarrow$1. We summarize these lifetimes in Table~\ref{table:LW}.

\section{Phonon-assisted indirect emission}

 In the optical limit $\mathrm{\textbf{Q}}\rightarrow0$ by solving and below the lowest bright exciton at 4.47 eV, we identified three additional pairs of degenerate, dark excitons (with $E$-type symmetry) located at 4.38 eV, 4.39 eV, and 4.40 eV, respectively. These direct excitons, due to their small oscillator strengths, are typically overshadowed in absorption spectra. Since the QP indirect gap in 2D h-AlN is smaller than the corresponding direct gap, therefore, understanding the emission spectra requires going beyond $\mathrm{\textbf{Q}}\rightarrow0$ limit to include excitons with finite (center of mass) momentum ($\mathrm{\textbf{Q}}\neq0$)~\cite{Bange_2023,Yadav_2023}. 
\begin{figure}[t]
  \centering
  \includegraphics[width=1\linewidth]{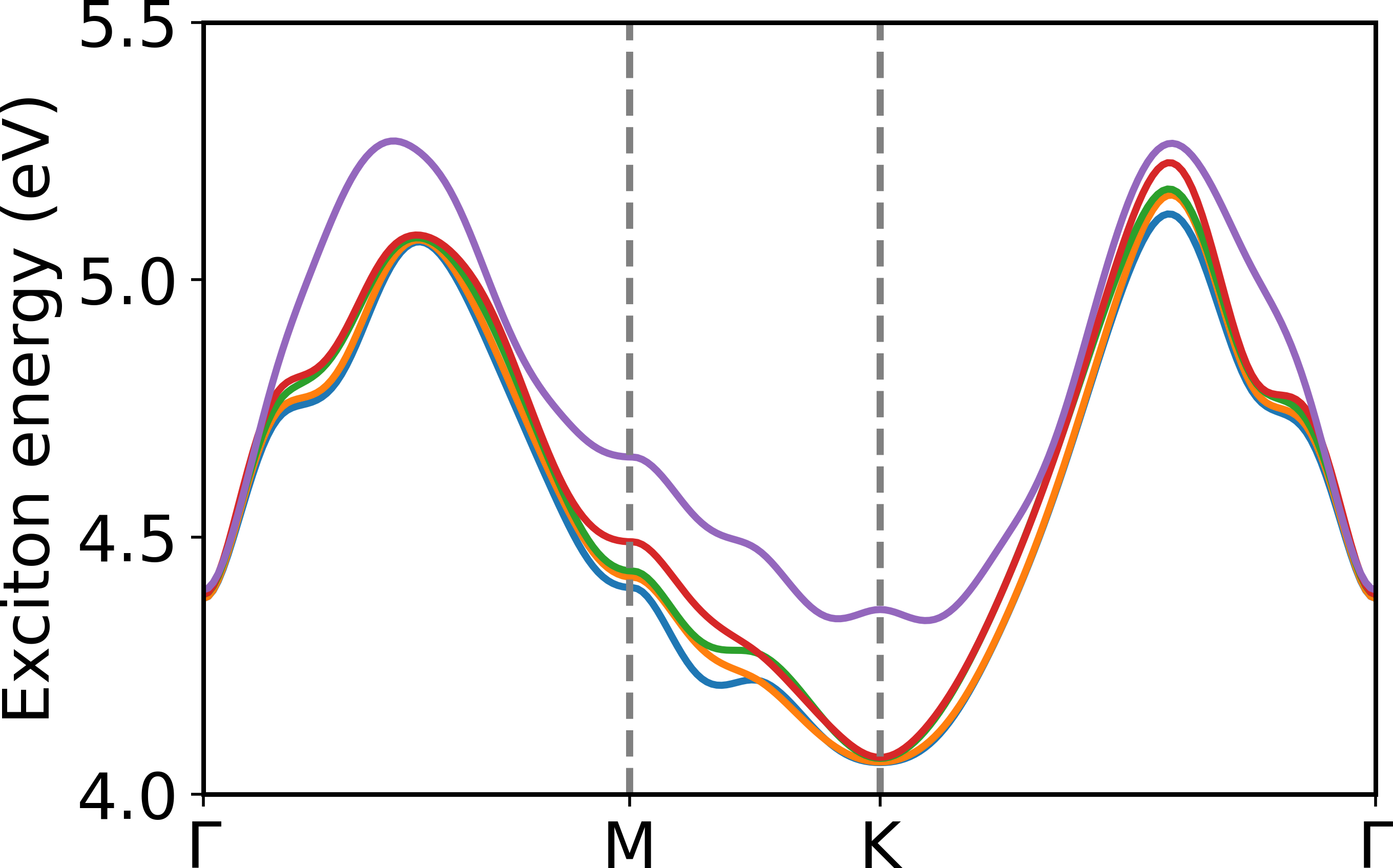}
  \caption{Finite momentum exciton dispersion in 2D AlN. The band-structure is shown for the lowest five excitons. The exciton energies at $\Gamma$ are the optical limit excitons.} 
  \label{fig5}
 \end{figure}
 
 \begin{figure*}[t]
  \centering
  \includegraphics[width=1\linewidth]{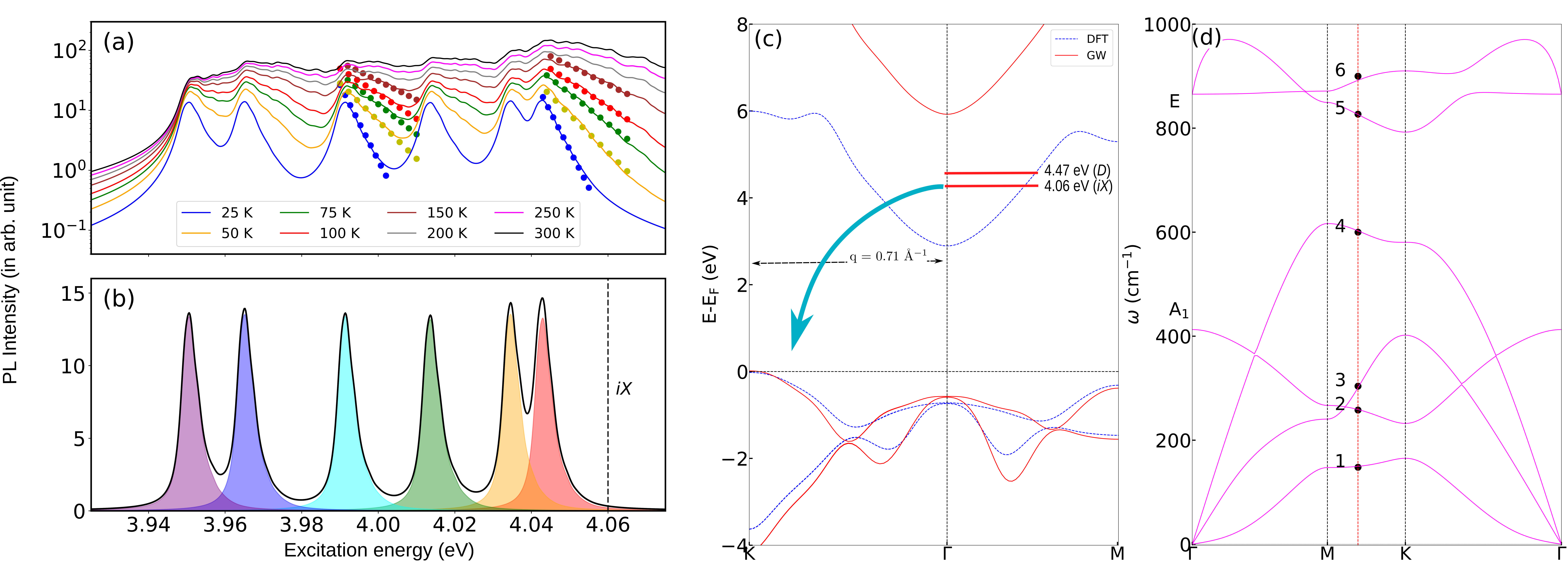}
  \caption{(a) PLE spectra at various temperature in 2D AlN. The sharp ripples at lower temperatures correspond to phonon replicas due to various modes. The dotted symbols correspond to the falling edge of the exponential dependency of the exciton thermalization. (b) PLE at 25 K resolving individual phonon mode assistance. The vertical line is the indirect exciton \textit{iX} located at 4.06 eV. (c) DFT (blue) and excited state (red) electronic band structure of 2D AlN. The blue (curved) arrow shows a schematic indirect electron-hole recombination assisted by phonons. The direct (D) and indirect (I) optical gaps are mentioned as levels. The top of the valence band in both cases is set to zero. (d) Corresponding lattice vibrations showing degenerate LO-TO mode at $\Gamma$. The black dots shows the phonon modes assisting the PLE process at finite transferred momentum, $\textbf{q}$ = 0.71 \AA$^{-1}$. Temperature-dependent PLE in 2D AlN highlights phonon-assisted exciton processes, with distinct phonon replicas at low temperatures, and reveals the role of indirect excitons and lattice vibrations in defining the optical response.} 
  \label{fig6}
 \end{figure*}

  The exciton bandstructure along high-symmetry path (for the exciton center of mass momentum) in the BZ is illustrated in figure~\ref{fig5}, capturing the dispersion of the lowest five excitons. It's worth noting that the exciton-transferred momentum $\Gamma$ represents the $\mathrm{\textbf{Q}}\rightarrow0$ excitonic energies demonstrated in the absorption spectrum. Along the $\Gamma$-\textbf{K} direction, we observe that the degeneracy in the exciton energies is lifted due to non $E$-type symmetry. The exciton dispersion reveals a minimum energy of 4.06 eV at the transferred momentum at \textbf{K} point ($|\mathrm{\textbf{Q}}| = 0.71$ \AA), reflecting an indirect optical gap resulting from a degenerate dark exciton pair ($i$X) below the direct bright exciton at 4.47 eV. These $i$X pairs are formed by an electron-hole pair coupled around the $\Gamma$ and \textbf{K} points of the electronic BZ (see figure~\ref{fig6}(c)). Due to the momentum conservation, this lowest $i$X pair shows weak oscillator strength. 
  
We calculate the energy difference between the lowest optically dark exciton pair at \textbf{K} and the first direct optically bright (4.47 eV) (4.38 eV) excitons at $\Gamma$, respectively. 
We find that the energy difference for the former is 0.41 eV, while for the latter case, it is 0.32 eV. Such differences are close to the crystal Debye energy ($\sim$0.12 eV), indicating assistance from optical phonons during recombination. 
We identify the phonon modes responsible for this assistance during emission (see figure~\ref{fig6}(c)). These are shown as dots in the phonon dispersion in figure~\ref{fig6}(d). These branches are between \textbf{M} and \textbf{K} points with phonon momentum of 0.71 $\AA$, required for the momentum conservation in the process. The lowest two branches [denoted by 1 and 2 in figure~\ref{fig6}(d)] correspond to out-of-plane acoustic vibrations, while the next higher branch (denoted by 3) represents an acoustic in-plane longitudinal motion. The mid-frequency (mode denoted by 4) corresponds to longitudinal optical in-plane vibrations, whereas the modes 5 and 6 correspond to in-plane circular vibrations, stretching and compressing the layer along longitudinal and transverse directions.

\begin{figure}[t]
  \centering
  \includegraphics[width=\linewidth]{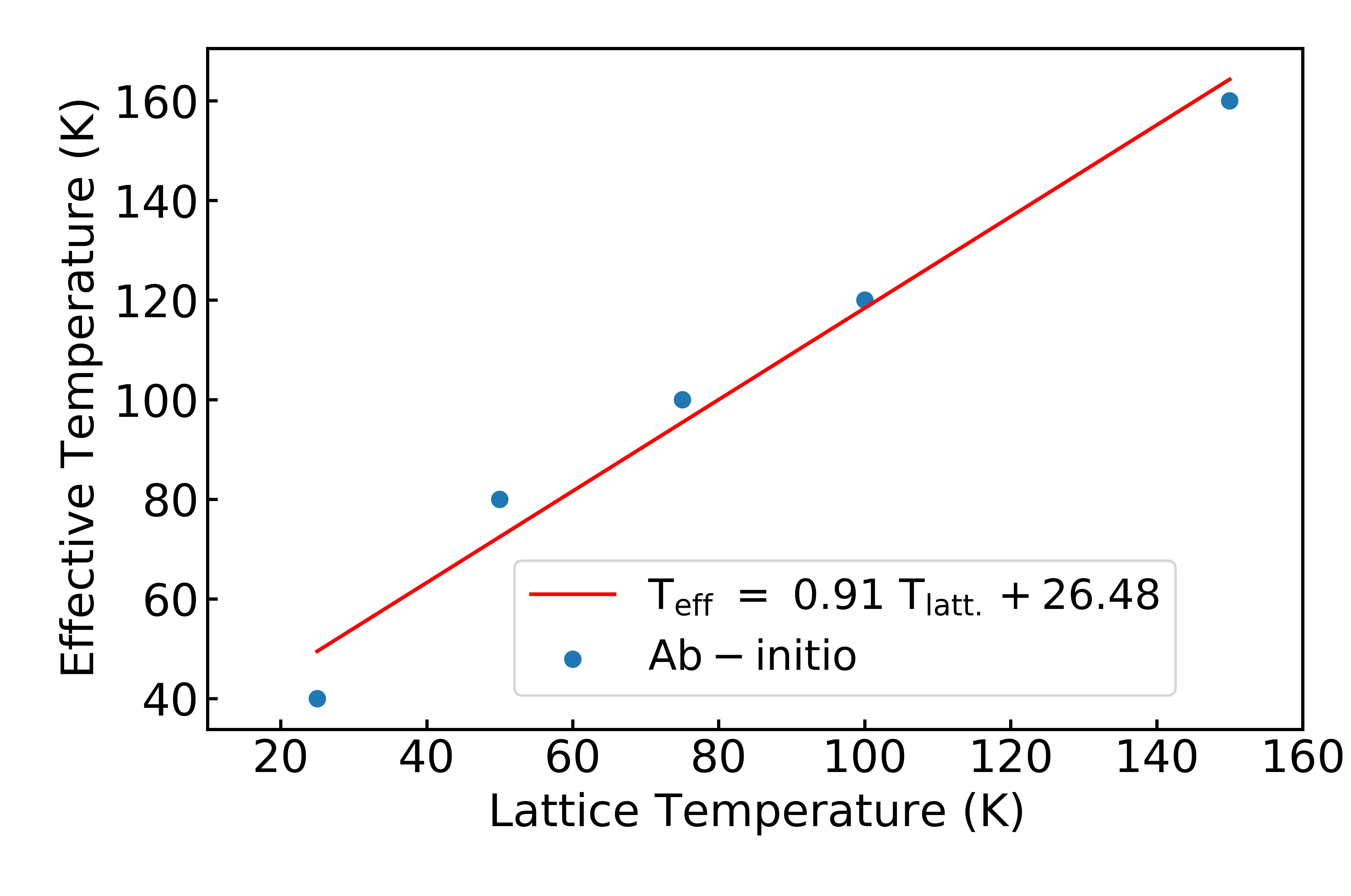}
  \caption{Exciton thermalization in 2D AlN. The dots correspond to the variation of the effective excitonic temperature with the lattice temperature. The straight line is the best fit with a slope 0.91. This highlights  that for $T>26$ K, the phonons play a significant role in the thermalization of the charge carriers.} 
  \label{fig7}
 \end{figure}

 In figure~\ref{fig6}(a)\ we illustrate the phonon-assisted DOS for PLE at various temperatures. Following Paleari ~\cite{Paleari2019}, the phonon-assisted DOS at frequency $\omega$ and temperature $T$ can be expressed as
\begin{equation}
\varrho=\sum_{\mathrm{s},Q,\lambda}\left[1+n_{B}\left(\omega_{\textbf{Q}}^{\lambda}\right)\right]\exp\left[-\frac{\mathcal{E}_{\textbf{Q}}^{\mathrm{s}}-\mathcal{E}_{min}}{k_{B}T}\right]\delta\left(\mathcal{E}_{\textbf{Q}}^{\mathrm{s}}-\omega_{\textbf{Q}}^{\lambda}-\omega\right)~.
\label{PL}
\end{equation}
Here, $\mathcal{E}_{\textbf{Q}}^{\mathrm{s}}$ is the energy of $\mathrm{s}^{th}$ exciton with momentum \textbf{Q} and $\mathcal{E}_{min}$ is the minimum excitonic energy. 
For our numerical calculations, we use a broadening of approximately 3 meV. 
This phonon-assisted spectra bypasses the rigorous consideration of exciton-phonon matrix elements and dipoles and the severe complexities associated with such calculations. Still, the luminescence spectra obtained using $\varrho\left(\omega,T\right)$ captures the phonon replicas associated with emission processes and generally matches well with experimental results~\cite{Cassabois2016}.  
In figure \ref{fig6}(a), we highlight several indirect emission processes. Notably, two distinct sets of bands near 4.02-4.05 and 3.94-4.00 eV emerge as the PLE lines. These lines exhibit a red-shift compared to the indirect exciton at 4.06 eV, indicating the necessity of phonon assistance for this emission process. Furthermore, examining the high-energy tails near the 4.00 and 4.05 eV lines, we discern an exponential fall-off rate, depicted by dotted symbols for the initial temperatures. From the PL spectrum, we observe that as the temperature rises, there is a reduction in the corresponding slope at the higher energy end. 
During the carrier relaxation process, the energy distribution of the excited carriers evolves to eventually align with a thermal Boltzmann distribution facilitated by phonon-assisted scatterings. To comprehend the thermalization of excitons, we employ a fitting approach akin to that used by Cassabois $et$ $al$. ~\cite{Cassabois2016}. 
In this approach, we use an exponential Boltzmann factor with an effective temperature $\mathrm{T}_{\mathrm{eff}}$ to fit the declining edge of the spectra. Our analysis in figure~\ref{fig7} shows a linear relationship between the effective temperature ($\mathrm{T}_{\mathrm{eff}}$) and at higher lattice temperatures ($\mathrm{T}_{\mathrm{lat}}$). The linear relation suggests that exciton thermalization with the crystal occurs at more than 26 K. This observation mirrors a recently reported scenario concerning emission spectra from monolayer h-BN ~\cite{Lechifflart2023}. 


Our detailed analysis of light emission processes from 2D h-AlN, provides critical insights for designing UV emitters. Understanding these phonon-assisted mechanisms and excitonic interactions lays the groundwork for optimizing the 2D AlN performance in optoelectronic applications in the UV regime. 

\section{Conclusions}
\label{conclusion}

Photoluminescence emissions provide critical insights into excitonic interactions, phonon coupling, and the effects of external perturbations, enhancing our understanding of the unique optical properties and potential applications of quantum materials in optoelectronic devices. We present a comprehensive exploration of the absorption and emission characteristics of two-dimensional hexagonal aluminum nitride (2D h-AlN). Our findings indicate that 2D h-AlN is an indirect semiconductor with a wide band gap of 5.73 eV, making it suitable for applications in the ultraviolet regime.

Accounting for many-body interactions such as electron-electron, electron-phonon, electron-hole, and exciton-phonon coupling, we accurately describe the excitonic spectrum of AlN, identifying excitons with binding energies up to 1.83 eV. Demonstrating strong electron-phonon correlation in AlN, we find that even in the low-temperature limit, the optical gap red-shifts significantly by 200 meV compared to scenarios neglecting lattice vibrations.
We show that non-radiative lifetimes for major bound excitons surpass their radiative lifetimes even at room temperature, with non-radiative processes occurring on a timescale of approximately 2.33 fs compared to radiative processes lasting around 770 fs. Extending beyond the optical dipole limit, our investigation into the excitonic dispersion reveals that photoluminescence emission in 2D h-AlN is a phonon-assisted process. A significant asymmetry emerges between the absorption and recombination channels, distinctly situating the two spectra around the indirect exciton gap. Our analysis identifies emission lines between 3.94-4.05 eV, marked by the presence of phonon replicas. Additionally, we show that excitonic thermalization becomes prominent above 26 K.

These exceptional optical properties of 2D h-AlN highlight its potential for developing efficient ultraviolet photon emitters.

\begin{acknowledgments}
\noindent 
This work was carried out with financial support from SERB, India with grant numbers CRG/2023/000476 and MTR/2021/000017.
P.Y. acknowledges UGC for the Senior Research Fellowship. We acknowledge the National Super Computing Mission (NSM) for providing computing resources of ``Paramsanganak'' at IIT Kanpur and ``Paramshivay'' at IIT BHU, India for the computational load. 
\end{acknowledgments}

\bibliography{new_ref}

\begin{thebibliography}{73}%
\makeatletter
\providecommand \@ifxundefined [1]{%
 \@ifx{#1\undefined}
}%
\providecommand \@ifnum [1]{%
 \ifnum #1\expandafter \@firstoftwo
 \else \expandafter \@secondoftwo
 \fi
}%
\providecommand \@ifx [1]{%
 \ifx #1\expandafter \@firstoftwo
 \else \expandafter \@secondoftwo
 \fi
}%
\providecommand \natexlab [1]{#1}%
\providecommand \enquote  [1]{``#1''}%
\providecommand \bibnamefont  [1]{#1}%
\providecommand \bibfnamefont [1]{#1}%
\providecommand \citenamefont [1]{#1}%
\providecommand \href@noop [0]{\@secondoftwo}%
\providecommand \href [0]{\begingroup \@sanitize@url \@href}%
\providecommand \@href[1]{\@@startlink{#1}\@@href}%
\providecommand \@@href[1]{\endgroup#1\@@endlink}%
\providecommand \@sanitize@url [0]{\catcode `\\12\catcode `\$12\catcode
  `\&12\catcode `\#12\catcode `\^12\catcode `\_12\catcode `\%12\relax}%
\providecommand \@@startlink[1]{}%
\providecommand \@@endlink[0]{}%
\providecommand \url  [0]{\begingroup\@sanitize@url \@url }%
\providecommand \@url [1]{\endgroup\@href {#1}{\urlprefix }}%
\providecommand \urlprefix  [0]{URL }%
\providecommand \Eprint [0]{\href }%
\providecommand \doibase [0]{https://doi.org/}%
\providecommand \selectlanguage [0]{\@gobble}%
\providecommand \bibinfo  [0]{\@secondoftwo}%
\providecommand \bibfield  [0]{\@secondoftwo}%
\providecommand \translation [1]{[#1]}%
\providecommand \BibitemOpen [0]{}%
\providecommand \bibitemStop [0]{}%
\providecommand \bibitemNoStop [0]{.\EOS\space}%
\providecommand \EOS [0]{\spacefactor3000\relax}%
\providecommand \BibitemShut  [1]{\csname bibitem#1\endcsname}%
\let\auto@bib@innerbib\@empty
\bibitem [{\citenamefont {Watanabe}\ \emph {et~al.}(2004)\citenamefont
  {Watanabe}, \citenamefont {Taniguchi},\ and\ \citenamefont
  {Kanda}}]{Watanabe2004}%
  \BibitemOpen
  \bibfield  {author} {\bibinfo {author} {\bibfnamefont {K.}~\bibnamefont
  {Watanabe}}, \bibinfo {author} {\bibfnamefont {T.}~\bibnamefont
  {Taniguchi}},\ and\ \bibinfo {author} {\bibfnamefont {H.}~\bibnamefont
  {Kanda}},\ }\bibfield  {title} {\bibinfo {title} {Direct-bandgap properties
  and evidence for ultraviolet lasing of hexagonal boron nitride single
  crystal},\ }\href {https://doi.org/10.1038/nmat1134} {\bibfield  {journal}
  {\bibinfo  {journal} {Nature Materials}\ }\textbf {\bibinfo {volume} {3}},\
  \bibinfo {pages} {404–409} (\bibinfo {year} {2004})}\BibitemShut {NoStop}%
\bibitem [{\citenamefont {Pozo-Zamudio}\ \emph {et~al.}(2015)\citenamefont
  {Pozo-Zamudio}, \citenamefont {Schwarz}, \citenamefont {Sich}, \citenamefont
  {Akimov}, \citenamefont {Bayer}, \citenamefont {Schofield}, \citenamefont
  {Chekhovich}, \citenamefont {Robinson}, \citenamefont {Kay}, \citenamefont
  {Kolosov}, \citenamefont {Dmitriev}, \citenamefont {Lashkarev}, \citenamefont
  {Borisenko}, \citenamefont {Kolesnikov},\ and\ \citenamefont
  {Tartakovskii}}]{Pozo-Zamudio_2015}%
  \BibitemOpen
  \bibfield  {author} {\bibinfo {author} {\bibfnamefont {O.~D.}\ \bibnamefont
  {Pozo-Zamudio}}, \bibinfo {author} {\bibfnamefont {S.}~\bibnamefont
  {Schwarz}}, \bibinfo {author} {\bibfnamefont {M.}~\bibnamefont {Sich}},
  \bibinfo {author} {\bibfnamefont {I.~A.}\ \bibnamefont {Akimov}}, \bibinfo
  {author} {\bibfnamefont {M.}~\bibnamefont {Bayer}}, \bibinfo {author}
  {\bibfnamefont {R.~C.}\ \bibnamefont {Schofield}}, \bibinfo {author}
  {\bibfnamefont {E.~A.}\ \bibnamefont {Chekhovich}}, \bibinfo {author}
  {\bibfnamefont {B.~J.}\ \bibnamefont {Robinson}}, \bibinfo {author}
  {\bibfnamefont {N.~D.}\ \bibnamefont {Kay}}, \bibinfo {author} {\bibfnamefont
  {O.~V.}\ \bibnamefont {Kolosov}}, \bibinfo {author} {\bibfnamefont {A.~I.}\
  \bibnamefont {Dmitriev}}, \bibinfo {author} {\bibfnamefont {G.~V.}\
  \bibnamefont {Lashkarev}}, \bibinfo {author} {\bibfnamefont {D.~N.}\
  \bibnamefont {Borisenko}}, \bibinfo {author} {\bibfnamefont {N.~N.}\
  \bibnamefont {Kolesnikov}},\ and\ \bibinfo {author} {\bibfnamefont {A.~I.}\
  \bibnamefont {Tartakovskii}},\ }\bibfield  {title} {\bibinfo {title}
  {Photoluminescence of two-dimensional gate and gase films},\ }\href
  {https://doi.org/10.1088/2053-1583/2/3/035010} {\bibfield  {journal}
  {\bibinfo  {journal} {2D Materials}\ }\textbf {\bibinfo {volume} {2}},\
  \bibinfo {pages} {035010} (\bibinfo {year} {2015})}\BibitemShut {NoStop}%
\bibitem [{\citenamefont {Cassabois}\ \emph {et~al.}(2016)\citenamefont
  {Cassabois}, \citenamefont {Valvin},\ and\ \citenamefont
  {Gil}}]{Cassabois2016}%
  \BibitemOpen
  \bibfield  {author} {\bibinfo {author} {\bibfnamefont {G.}~\bibnamefont
  {Cassabois}}, \bibinfo {author} {\bibfnamefont {P.}~\bibnamefont {Valvin}},\
  and\ \bibinfo {author} {\bibfnamefont {B.}~\bibnamefont {Gil}},\ }\bibfield
  {title} {\bibinfo {title} {Hexagonal boron nitride is an indirect bandgap
  semiconductor},\ }\href {https://doi.org/10.1038/nphoton.2015.277} {\bibfield
   {journal} {\bibinfo  {journal} {Nature Photonics}\ }\textbf {\bibinfo
  {volume} {10}},\ \bibinfo {pages} {262} (\bibinfo {year} {2016})}\BibitemShut
  {NoStop}%
\bibitem [{\citenamefont {Molas}\ \emph {et~al.}(2017)\citenamefont {Molas},
  \citenamefont {Faugeras}, \citenamefont {Slobodeniuk}, \citenamefont
  {Nogajewski}, \citenamefont {Bartos}, \citenamefont {Basko},\ and\
  \citenamefont {Potemski}}]{Molas2017}%
  \BibitemOpen
  \bibfield  {author} {\bibinfo {author} {\bibfnamefont {M.~R.}\ \bibnamefont
  {Molas}}, \bibinfo {author} {\bibfnamefont {C.}~\bibnamefont {Faugeras}},
  \bibinfo {author} {\bibfnamefont {A.~O.}\ \bibnamefont {Slobodeniuk}},
  \bibinfo {author} {\bibfnamefont {K.}~\bibnamefont {Nogajewski}}, \bibinfo
  {author} {\bibfnamefont {M.}~\bibnamefont {Bartos}}, \bibinfo {author}
  {\bibfnamefont {D.~M.}\ \bibnamefont {Basko}},\ and\ \bibinfo {author}
  {\bibfnamefont {M.}~\bibnamefont {Potemski}},\ }\bibfield  {title} {\bibinfo
  {title} {Brightening of dark excitons in monolayers of semiconducting
  transition metal dichalcogenides},\ }\href
  {https://doi.org/10.1088/2053-1583/aa5521} {\bibfield  {journal} {\bibinfo
  {journal} {2D Materials}\ }\textbf {\bibinfo {volume} {4}},\ \bibinfo {pages}
  {021003} (\bibinfo {year} {2017})}\BibitemShut {NoStop}%
\bibitem [{\citenamefont {Karmakar}\ \emph {et~al.}(2021)\citenamefont
  {Karmakar}, \citenamefont {Bhattacharya}, \citenamefont {Mukherjee},
  \citenamefont {Ghosh}, \citenamefont {Chowdhury}, \citenamefont {Agarwal},
  \citenamefont {Ray}, \citenamefont {Chanda},\ and\ \citenamefont
  {Datta}}]{Karmakar2021}%
  \BibitemOpen
  \bibfield  {author} {\bibinfo {author} {\bibfnamefont {M.}~\bibnamefont
  {Karmakar}}, \bibinfo {author} {\bibfnamefont {S.}~\bibnamefont
  {Bhattacharya}}, \bibinfo {author} {\bibfnamefont {S.}~\bibnamefont
  {Mukherjee}}, \bibinfo {author} {\bibfnamefont {B.}~\bibnamefont {Ghosh}},
  \bibinfo {author} {\bibfnamefont {R.~K.}\ \bibnamefont {Chowdhury}}, \bibinfo
  {author} {\bibfnamefont {A.}~\bibnamefont {Agarwal}}, \bibinfo {author}
  {\bibfnamefont {S.~K.}\ \bibnamefont {Ray}}, \bibinfo {author} {\bibfnamefont
  {D.}~\bibnamefont {Chanda}},\ and\ \bibinfo {author} {\bibfnamefont {P.~K.}\
  \bibnamefont {Datta}},\ }\bibfield  {title} {\bibinfo {title} {Observation of
  dynamic screening in the excited exciton states in multilayered
  ${\mathrm{mos}}_{2}$},\ }\href {https://doi.org/10.1103/PhysRevB.103.075437}
  {\bibfield  {journal} {\bibinfo  {journal} {Phys. Rev. B}\ }\textbf {\bibinfo
  {volume} {103}},\ \bibinfo {pages} {075437} (\bibinfo {year}
  {2021})}\BibitemShut {NoStop}%
\bibitem [{\citenamefont {Thygesen}(2017)}]{Thygesen2017}%
  \BibitemOpen
  \bibfield  {author} {\bibinfo {author} {\bibfnamefont {K.~S.}\ \bibnamefont
  {Thygesen}},\ }\bibfield  {title} {\bibinfo {title} {Calculating excitons,
  plasmons, and quasiparticles in 2d materials and van der waals
  heterostructures},\ }\href {https://doi.org/10.1088/2053-1583/aa6432}
  {\bibfield  {journal} {\bibinfo  {journal} {2D Mater.}\ }\textbf {\bibinfo
  {volume} {4}},\ \bibinfo {pages} {022004} (\bibinfo {year}
  {2017})}\BibitemShut {NoStop}%
\bibitem [{\citenamefont {Mounet}\ \emph {et~al.}(2018)\citenamefont {Mounet},
  \citenamefont {Gibertini}, \citenamefont {Schwaller}, \citenamefont {Campi},
  \citenamefont {Merkys}, \citenamefont {Marrazzo}, \citenamefont {Sohier},
  \citenamefont {Castelli}, \citenamefont {Cepellotti}, \citenamefont {Pizzi},\
  and\ \citenamefont {Marzari}}]{Mounet2018}%
  \BibitemOpen
  \bibfield  {author} {\bibinfo {author} {\bibfnamefont {N.}~\bibnamefont
  {Mounet}}, \bibinfo {author} {\bibfnamefont {M.}~\bibnamefont {Gibertini}},
  \bibinfo {author} {\bibfnamefont {P.}~\bibnamefont {Schwaller}}, \bibinfo
  {author} {\bibfnamefont {D.}~\bibnamefont {Campi}}, \bibinfo {author}
  {\bibfnamefont {A.}~\bibnamefont {Merkys}}, \bibinfo {author} {\bibfnamefont
  {A.}~\bibnamefont {Marrazzo}}, \bibinfo {author} {\bibfnamefont
  {T.}~\bibnamefont {Sohier}}, \bibinfo {author} {\bibfnamefont {I.~E.}\
  \bibnamefont {Castelli}}, \bibinfo {author} {\bibfnamefont {A.}~\bibnamefont
  {Cepellotti}}, \bibinfo {author} {\bibfnamefont {G.}~\bibnamefont {Pizzi}},\
  and\ \bibinfo {author} {\bibfnamefont {N.}~\bibnamefont {Marzari}},\
  }\bibfield  {title} {\bibinfo {title} {Two-dimensional materials from
  high-throughput computational exfoliation of experimentally known
  compounds},\ }\href {https://doi.org/10.1038/s41565-017-0035-5} {\bibfield
  {journal} {\bibinfo  {journal} {Nature Nanotechnology}\ }\textbf {\bibinfo
  {volume} {13}},\ \bibinfo {pages} {246} (\bibinfo {year} {2018})}\BibitemShut
  {NoStop}%
\bibitem [{\citenamefont {Bera}\ \emph {et~al.}(2021)\citenamefont {Bera},
  \citenamefont {Shrivastava}, \citenamefont {Bramhachari}, \citenamefont
  {Zhang}, \citenamefont {Poonia}, \citenamefont {Mandal}, \citenamefont
  {Miller}, \citenamefont {Beard}, \citenamefont {Agarwal},\ and\ \citenamefont
  {Adarsh}}]{Santu2021}%
  \BibitemOpen
  \bibfield  {author} {\bibinfo {author} {\bibfnamefont {S.~K.}\ \bibnamefont
  {Bera}}, \bibinfo {author} {\bibfnamefont {M.}~\bibnamefont {Shrivastava}},
  \bibinfo {author} {\bibfnamefont {K.}~\bibnamefont {Bramhachari}}, \bibinfo
  {author} {\bibfnamefont {H.}~\bibnamefont {Zhang}}, \bibinfo {author}
  {\bibfnamefont {A.~K.}\ \bibnamefont {Poonia}}, \bibinfo {author}
  {\bibfnamefont {D.}~\bibnamefont {Mandal}}, \bibinfo {author} {\bibfnamefont
  {E.~M.}\ \bibnamefont {Miller}}, \bibinfo {author} {\bibfnamefont {M.~C.}\
  \bibnamefont {Beard}}, \bibinfo {author} {\bibfnamefont {A.}~\bibnamefont
  {Agarwal}},\ and\ \bibinfo {author} {\bibfnamefont {K.~V.}\ \bibnamefont
  {Adarsh}},\ }\bibfield  {title} {\bibinfo {title} {Atomlike interaction and
  optically tunable giant band-gap renormalization in large-area atomically
  thin ${\mathrm{mos}}_{2}$},\ }\href
  {https://doi.org/10.1103/PhysRevB.104.L201404} {\bibfield  {journal}
  {\bibinfo  {journal} {Phys. Rev. B}\ }\textbf {\bibinfo {volume} {104}},\
  \bibinfo {pages} {L201404} (\bibinfo {year} {2021})}\BibitemShut {NoStop}%
\bibitem [{\citenamefont {Yadav}\ \emph
  {et~al.}(2023{\natexlab{a}})\citenamefont {Yadav}, \citenamefont {Adarsh},\
  and\ \citenamefont {Agarwal}}]{Yadav_2023_EHL-2D}%
  \BibitemOpen
  \bibfield  {author} {\bibinfo {author} {\bibfnamefont {P.}~\bibnamefont
  {Yadav}}, \bibinfo {author} {\bibfnamefont {K.~V.}\ \bibnamefont {Adarsh}},\
  and\ \bibinfo {author} {\bibfnamefont {A.}~\bibnamefont {Agarwal}},\
  }\bibfield  {title} {\bibinfo {title} {Room temperature electron–hole
  liquid phase in monolayer mosi2z4 (z = pinctogen)},\ }\href
  {https://doi.org/10.1088/2053-1583/ace83b} {\bibfield  {journal} {\bibinfo
  {journal} {2D Materials}\ }\textbf {\bibinfo {volume} {10}},\ \bibinfo
  {pages} {045007} (\bibinfo {year} {2023}{\natexlab{a}})}\BibitemShut
  {NoStop}%
\bibitem [{\citenamefont {Wang}\ \emph {et~al.}(2014)\citenamefont {Wang},
  \citenamefont {Bouet}, \citenamefont {Lagarde}, \citenamefont {Vidal},
  \citenamefont {Balocchi}, \citenamefont {Amand}, \citenamefont {Marie},\ and\
  \citenamefont {Urbaszek}}]{Wang2014PL}%
  \BibitemOpen
  \bibfield  {author} {\bibinfo {author} {\bibfnamefont {G.}~\bibnamefont
  {Wang}}, \bibinfo {author} {\bibfnamefont {L.}~\bibnamefont {Bouet}},
  \bibinfo {author} {\bibfnamefont {D.}~\bibnamefont {Lagarde}}, \bibinfo
  {author} {\bibfnamefont {M.}~\bibnamefont {Vidal}}, \bibinfo {author}
  {\bibfnamefont {A.}~\bibnamefont {Balocchi}}, \bibinfo {author}
  {\bibfnamefont {T.}~\bibnamefont {Amand}}, \bibinfo {author} {\bibfnamefont
  {X.}~\bibnamefont {Marie}},\ and\ \bibinfo {author} {\bibfnamefont
  {B.}~\bibnamefont {Urbaszek}},\ }\bibfield  {title} {\bibinfo {title} {Valley
  dynamics probed through charged and neutral exciton emission in monolayer
  ${\mathrm{wse}}_{2}$},\ }\href {https://doi.org/10.1103/PhysRevB.90.075413}
  {\bibfield  {journal} {\bibinfo  {journal} {Phys. Rev. B}\ }\textbf {\bibinfo
  {volume} {90}},\ \bibinfo {pages} {075413} (\bibinfo {year}
  {2014})}\BibitemShut {NoStop}%
\bibitem [{\citenamefont {Brem}\ \emph {et~al.}(2020)\citenamefont {Brem},
  \citenamefont {Ekman}, \citenamefont {Christiansen}, \citenamefont {Katsch},
  \citenamefont {Selig}, \citenamefont {Robert}, \citenamefont {Marie},
  \citenamefont {Urbaszek}, \citenamefont {Knorr},\ and\ \citenamefont
  {Malic}}]{Brem2020}%
  \BibitemOpen
  \bibfield  {author} {\bibinfo {author} {\bibfnamefont {S.}~\bibnamefont
  {Brem}}, \bibinfo {author} {\bibfnamefont {A.}~\bibnamefont {Ekman}},
  \bibinfo {author} {\bibfnamefont {D.}~\bibnamefont {Christiansen}}, \bibinfo
  {author} {\bibfnamefont {F.}~\bibnamefont {Katsch}}, \bibinfo {author}
  {\bibfnamefont {M.}~\bibnamefont {Selig}}, \bibinfo {author} {\bibfnamefont
  {C.}~\bibnamefont {Robert}}, \bibinfo {author} {\bibfnamefont
  {X.}~\bibnamefont {Marie}}, \bibinfo {author} {\bibfnamefont
  {B.}~\bibnamefont {Urbaszek}}, \bibinfo {author} {\bibfnamefont
  {A.}~\bibnamefont {Knorr}},\ and\ \bibinfo {author} {\bibfnamefont
  {E.}~\bibnamefont {Malic}},\ }\bibfield  {title} {\bibinfo {title}
  {Phonon-assisted photoluminescence from indirect excitons in monolayers of
  transition-metal dichalcogenides},\ }\href
  {https://doi.org/10.1021/acs.nanolett.0c00633} {\bibfield  {journal}
  {\bibinfo  {journal} {Nano Letters}\ }\textbf {\bibinfo {volume} {20}},\
  \bibinfo {pages} {2849} (\bibinfo {year} {2020})}\BibitemShut {NoStop}%
\bibitem [{\citenamefont {Marini}(2008)}]{Marini2008}%
  \BibitemOpen
  \bibfield  {author} {\bibinfo {author} {\bibfnamefont {A.}~\bibnamefont
  {Marini}},\ }\bibfield  {title} {\bibinfo {title} {Ab initio
  finite-temperature excitons},\ }\href
  {https://doi.org/10.1103/PhysRevLett.101.106405} {\bibfield  {journal}
  {\bibinfo  {journal} {Phys. Rev. Lett.}\ }\textbf {\bibinfo {volume} {101}},\
  \bibinfo {pages} {106405} (\bibinfo {year} {2008})}\BibitemShut {NoStop}%
\bibitem [{\citenamefont {Hedin}(1965)}]{HedinGW1}%
  \BibitemOpen
  \bibfield  {author} {\bibinfo {author} {\bibfnamefont {L.}~\bibnamefont
  {Hedin}},\ }\bibfield  {title} {\bibinfo {title} {New method for calculating
  the one-particle green's function with application to the electron-gas
  problem},\ }\href {https://doi.org/10.1103/PhysRev.139.A796} {\bibfield
  {journal} {\bibinfo  {journal} {Phys. Rev.}\ }\textbf {\bibinfo {volume}
  {139}},\ \bibinfo {pages} {A796} (\bibinfo {year} {1965})}\BibitemShut
  {NoStop}%
\bibitem [{\citenamefont {Hybertsen}\ and\ \citenamefont
  {Louie}(1986)}]{Hybertsen-GW2}%
  \BibitemOpen
  \bibfield  {author} {\bibinfo {author} {\bibfnamefont {M.~S.}\ \bibnamefont
  {Hybertsen}}\ and\ \bibinfo {author} {\bibfnamefont {S.~G.}\ \bibnamefont
  {Louie}},\ }\bibfield  {title} {\bibinfo {title} {Electron correlation in
  semiconductors and insulators: Band gaps and quasiparticle energies},\ }\href
  {https://doi.org/10.1103/PhysRevB.34.5390} {\bibfield  {journal} {\bibinfo
  {journal} {Phys. Rev. B}\ }\textbf {\bibinfo {volume} {34}},\ \bibinfo
  {pages} {5390} (\bibinfo {year} {1986})}\BibitemShut {NoStop}%
\bibitem [{\citenamefont {Noffsinger}\ \emph {et~al.}(2012)\citenamefont
  {Noffsinger}, \citenamefont {Kioupakis}, \citenamefont {Van~de Walle},
  \citenamefont {Louie},\ and\ \citenamefont {Cohen}}]{Noffsinger2012}%
  \BibitemOpen
  \bibfield  {author} {\bibinfo {author} {\bibfnamefont {J.}~\bibnamefont
  {Noffsinger}}, \bibinfo {author} {\bibfnamefont {E.}~\bibnamefont
  {Kioupakis}}, \bibinfo {author} {\bibfnamefont {C.~G.}\ \bibnamefont {Van~de
  Walle}}, \bibinfo {author} {\bibfnamefont {S.~G.}\ \bibnamefont {Louie}},\
  and\ \bibinfo {author} {\bibfnamefont {M.~L.}\ \bibnamefont {Cohen}},\
  }\bibfield  {title} {\bibinfo {title} {Phonon-assisted optical absorption in
  silicon from first principles},\ }\href
  {https://doi.org/10.1103/PhysRevLett.108.167402} {\bibfield  {journal}
  {\bibinfo  {journal} {Phys. Rev. Lett.}\ }\textbf {\bibinfo {volume} {108}},\
  \bibinfo {pages} {167402} (\bibinfo {year} {2012})}\BibitemShut {NoStop}%
\bibitem [{\citenamefont {Yang}\ and\ \citenamefont {Draxl}(2022)}]{Mao2022}%
  \BibitemOpen
  \bibfield  {author} {\bibinfo {author} {\bibfnamefont {M.}~\bibnamefont
  {Yang}}\ and\ \bibinfo {author} {\bibfnamefont {C.}~\bibnamefont {Draxl}},\
  }\href {https://arxiv.org/abs/2212.13645} {\bibinfo {title} {Novel approach
  to structural relaxation of materials in optically excited states}} (\bibinfo
  {year} {2022}),\ \Eprint {https://arxiv.org/abs/2212.13645} {arXiv:2212.13645
  [physics.comp-ph]} \BibitemShut {NoStop}%
\bibitem [{\citenamefont {Xiao}\ \emph {et~al.}(2012)\citenamefont {Xiao},
  \citenamefont {Liu}, \citenamefont {Feng}, \citenamefont {Xu},\ and\
  \citenamefont {Yao}}]{SOC-effect-MoS2-PRL}%
  \BibitemOpen
  \bibfield  {author} {\bibinfo {author} {\bibfnamefont {D.}~\bibnamefont
  {Xiao}}, \bibinfo {author} {\bibfnamefont {G.-B.}\ \bibnamefont {Liu}},
  \bibinfo {author} {\bibfnamefont {W.}~\bibnamefont {Feng}}, \bibinfo {author}
  {\bibfnamefont {X.}~\bibnamefont {Xu}},\ and\ \bibinfo {author}
  {\bibfnamefont {W.}~\bibnamefont {Yao}},\ }\bibfield  {title} {\bibinfo
  {title} {Coupled spin and valley physics in monolayers of
  ${\mathrm{mos}}_{2}$ and other group-vi dichalcogenides},\ }\href
  {https://doi.org/10.1103/PhysRevLett.108.196802} {\bibfield  {journal}
  {\bibinfo  {journal} {Phys. Rev. Lett.}\ }\textbf {\bibinfo {volume} {108}},\
  \bibinfo {pages} {196802} (\bibinfo {year} {2012})}\BibitemShut {NoStop}%
\bibitem [{\citenamefont {Wang}\ \emph {et~al.}(2018)\citenamefont {Wang},
  \citenamefont {Chernikov}, \citenamefont {Glazov}, \citenamefont {Heinz},
  \citenamefont {Marie}, \citenamefont {Amand},\ and\ \citenamefont
  {Urbaszek}}]{Wang2018}%
  \BibitemOpen
  \bibfield  {author} {\bibinfo {author} {\bibfnamefont {G.}~\bibnamefont
  {Wang}}, \bibinfo {author} {\bibfnamefont {A.}~\bibnamefont {Chernikov}},
  \bibinfo {author} {\bibfnamefont {M.~M.}\ \bibnamefont {Glazov}}, \bibinfo
  {author} {\bibfnamefont {T.~F.}\ \bibnamefont {Heinz}}, \bibinfo {author}
  {\bibfnamefont {X.}~\bibnamefont {Marie}}, \bibinfo {author} {\bibfnamefont
  {T.}~\bibnamefont {Amand}},\ and\ \bibinfo {author} {\bibfnamefont
  {B.}~\bibnamefont {Urbaszek}},\ }\bibfield  {title} {\bibinfo {title}
  {Colloquium: Excitons in atomically thin transition metal dichalcogenides},\
  }\href {https://doi.org/10.1103/RevModPhys.90.021001} {\bibfield  {journal}
  {\bibinfo  {journal} {Rev. Mod. Phys.}\ }\textbf {\bibinfo {volume} {90}},\
  \bibinfo {pages} {021001} (\bibinfo {year} {2018})}\BibitemShut {NoStop}%
\bibitem [{\citenamefont {Yadav}\ \emph
  {et~al.}(2023{\natexlab{b}})\citenamefont {Yadav}, \citenamefont {Khamari},
  \citenamefont {Singh}, \citenamefont {Adarsh},\ and\ \citenamefont
  {Agarwal}}]{Yadav_2023}%
  \BibitemOpen
  \bibfield  {author} {\bibinfo {author} {\bibfnamefont {P.}~\bibnamefont
  {Yadav}}, \bibinfo {author} {\bibfnamefont {B.}~\bibnamefont {Khamari}},
  \bibinfo {author} {\bibfnamefont {B.}~\bibnamefont {Singh}}, \bibinfo
  {author} {\bibfnamefont {K.~V.}\ \bibnamefont {Adarsh}},\ and\ \bibinfo
  {author} {\bibfnamefont {A.}~\bibnamefont {Agarwal}},\ }\bibfield  {title}
  {\bibinfo {title} {Fluence dependent dynamics of excitons in monolayer
  mosi2z4 (z = pnictogen)},\ }\href {https://doi.org/10.1088/1361-648X/acc43f}
  {\bibfield  {journal} {\bibinfo  {journal} {Journal of Physics: Condensed
  Matter}\ }\textbf {\bibinfo {volume} {35}},\ \bibinfo {pages} {235701}
  (\bibinfo {year} {2023}{\natexlab{b}})}\BibitemShut {NoStop}%
\bibitem [{\citenamefont {Zhuang}\ \emph
  {et~al.}(2013{\natexlab{a}})\citenamefont {Zhuang}, \citenamefont {Singh},\
  and\ \citenamefont {Hennig}}]{Zhuang13}%
  \BibitemOpen
  \bibfield  {author} {\bibinfo {author} {\bibfnamefont {H.~L.}\ \bibnamefont
  {Zhuang}}, \bibinfo {author} {\bibfnamefont {A.~K.}\ \bibnamefont {Singh}},\
  and\ \bibinfo {author} {\bibfnamefont {R.~G.}\ \bibnamefont {Hennig}},\
  }\bibfield  {title} {\bibinfo {title} {Computational discovery of
  single-layer iii-v materials},\ }\href
  {https://doi.org/10.1103/PhysRevB.87.165415} {\bibfield  {journal} {\bibinfo
  {journal} {Phys. Rev. B}\ }\textbf {\bibinfo {volume} {87}},\ \bibinfo
  {pages} {165415} (\bibinfo {year} {2013}{\natexlab{a}})}\BibitemShut
  {NoStop}%
\bibitem [{\citenamefont {Yadav}\ \emph {et~al.}(2015)\citenamefont {Yadav},
  \citenamefont {Duarte}, \citenamefont {Khandelwal}, \citenamefont {Agarwal},
  \citenamefont {Hu},\ and\ \citenamefont {Chauhan}}]{Chandan2015}%
  \BibitemOpen
  \bibfield  {author} {\bibinfo {author} {\bibfnamefont {C.}~\bibnamefont
  {Yadav}}, \bibinfo {author} {\bibfnamefont {J.~P.}\ \bibnamefont {Duarte}},
  \bibinfo {author} {\bibfnamefont {S.}~\bibnamefont {Khandelwal}}, \bibinfo
  {author} {\bibfnamefont {A.}~\bibnamefont {Agarwal}}, \bibinfo {author}
  {\bibfnamefont {C.}~\bibnamefont {Hu}},\ and\ \bibinfo {author}
  {\bibfnamefont {Y.~S.}\ \bibnamefont {Chauhan}},\ }\bibfield  {title}
  {\bibinfo {title} {Capacitance modeling in iii–v finfets},\ }\href
  {https://doi.org/10.1109/TED.2015.2480380} {\bibfield  {journal} {\bibinfo
  {journal} {IEEE Transactions on Electron Devices}\ }\textbf {\bibinfo
  {volume} {62}},\ \bibinfo {pages} {3892} (\bibinfo {year}
  {2015})}\BibitemShut {NoStop}%
\bibitem [{\citenamefont {Dutta}\ \emph {et~al.}(2016)\citenamefont {Dutta},
  \citenamefont {Kumar}, \citenamefont {Rastogi}, \citenamefont {Agarwal},\
  and\ \citenamefont {Chauhan}}]{Dutta2016}%
  \BibitemOpen
  \bibfield  {author} {\bibinfo {author} {\bibfnamefont {T.}~\bibnamefont
  {Dutta}}, \bibinfo {author} {\bibfnamefont {S.}~\bibnamefont {Kumar}},
  \bibinfo {author} {\bibfnamefont {P.}~\bibnamefont {Rastogi}}, \bibinfo
  {author} {\bibfnamefont {A.}~\bibnamefont {Agarwal}},\ and\ \bibinfo {author}
  {\bibfnamefont {Y.~S.}\ \bibnamefont {Chauhan}},\ }\bibfield  {title}
  {\bibinfo {title} {Impact of channel thickness variation on bandstructure and
  source-to-drain tunneling in ultra-thin body iii-v mosfets},\ }\href
  {https://doi.org/10.1109/JEDS.2016.2522981} {\bibfield  {journal} {\bibinfo
  {journal} {IEEE Journal of the Electron Devices Society}\ }\textbf {\bibinfo
  {volume} {4}},\ \bibinfo {pages} {66} (\bibinfo {year} {2016})}\BibitemShut
  {NoStop}%
\bibitem [{\citenamefont {Wu}\ \emph {et~al.}(2017)\citenamefont {Wu},
  \citenamefont {Yang}, \citenamefont {Gao}, \citenamefont {Qi}, \citenamefont
  {Zhang}, \citenamefont {Qiao},\ and\ \citenamefont {Ren}}]{Jiongyao2017}%
  \BibitemOpen
  \bibfield  {author} {\bibinfo {author} {\bibfnamefont {J.}~\bibnamefont
  {Wu}}, \bibinfo {author} {\bibfnamefont {Y.}~\bibnamefont {Yang}}, \bibinfo
  {author} {\bibfnamefont {H.}~\bibnamefont {Gao}}, \bibinfo {author}
  {\bibfnamefont {Y.}~\bibnamefont {Qi}}, \bibinfo {author} {\bibfnamefont
  {J.}~\bibnamefont {Zhang}}, \bibinfo {author} {\bibfnamefont
  {Z.}~\bibnamefont {Qiao}},\ and\ \bibinfo {author} {\bibfnamefont
  {W.}~\bibnamefont {Ren}},\ }\bibfield  {title} {\bibinfo {title} {Electric
  field effect of gaas monolayer from first principles},\ }\href
  {https://doi.org/https://doi.org/10.1063/1.4979507} {\bibfield  {journal}
  {\bibinfo  {journal} {AIP Adv.}\ }\textbf {\bibinfo {volume} {7}},\ \bibinfo
  {pages} {035218} (\bibinfo {year} {2017})}\BibitemShut {NoStop}%
\bibitem [{\citenamefont {Lucking}\ \emph {et~al.}(2018)\citenamefont
  {Lucking}, \citenamefont {Xie}, \citenamefont {Choe}, \citenamefont {West},
  \citenamefont {Lu},\ and\ \citenamefont {Zhang}}]{Michael2018}%
  \BibitemOpen
  \bibfield  {author} {\bibinfo {author} {\bibfnamefont {M.~C.}\ \bibnamefont
  {Lucking}}, \bibinfo {author} {\bibfnamefont {W.}~\bibnamefont {Xie}},
  \bibinfo {author} {\bibfnamefont {D.-H.}\ \bibnamefont {Choe}}, \bibinfo
  {author} {\bibfnamefont {D.}~\bibnamefont {West}}, \bibinfo {author}
  {\bibfnamefont {T.-M.}\ \bibnamefont {Lu}},\ and\ \bibinfo {author}
  {\bibfnamefont {S.~B.}\ \bibnamefont {Zhang}},\ }\bibfield  {title} {\bibinfo
  {title} {Traditional semiconductors in the two-dimensional limit},\ }\href
  {https://doi.org/https://doi.org/10.1103/PhysRevLett.120.086101} {\bibfield
  {journal} {\bibinfo  {journal} {Phys. Rev. Lett.}\ }\textbf {\bibinfo
  {volume} {120}},\ \bibinfo {pages} {086101} (\bibinfo {year}
  {2018})}\BibitemShut {NoStop}%
\bibitem [{\citenamefont {Jiang}\ \emph {et~al.}(2017)\citenamefont {Jiang},
  \citenamefont {Jing}, \citenamefont {Huang}, \citenamefont {Liu},
  \citenamefont {Du}, \citenamefont {Liu}, \citenamefont {Pu}, \citenamefont
  {Hu},\ and\ \citenamefont {Wang}}]{Jiang2017-GaN}%
  \BibitemOpen
  \bibfield  {author} {\bibinfo {author} {\bibfnamefont {C.}~\bibnamefont
  {Jiang}}, \bibinfo {author} {\bibfnamefont {L.}~\bibnamefont {Jing}},
  \bibinfo {author} {\bibfnamefont {X.}~\bibnamefont {Huang}}, \bibinfo
  {author} {\bibfnamefont {M.}~\bibnamefont {Liu}}, \bibinfo {author}
  {\bibfnamefont {C.}~\bibnamefont {Du}}, \bibinfo {author} {\bibfnamefont
  {T.}~\bibnamefont {Liu}}, \bibinfo {author} {\bibfnamefont {X.}~\bibnamefont
  {Pu}}, \bibinfo {author} {\bibfnamefont {W.}~\bibnamefont {Hu}},\ and\
  \bibinfo {author} {\bibfnamefont {Z.~L.}\ \bibnamefont {Wang}},\ }\bibfield
  {title} {\bibinfo {title} {Enhanced solar cell conversion efficiency of
  {InGaN}/{GaN} multiple quantum wells by piezo-phototronic effect},\ }\href
  {https://doi.org/10.1021/acsnano.7b04935} {\bibfield  {journal} {\bibinfo
  {journal} {{ACS} Nano}\ }\textbf {\bibinfo {volume} {11}},\ \bibinfo {pages}
  {9405} (\bibinfo {year} {2017})}\BibitemShut {NoStop}%
\bibitem [{\citenamefont {Attaccalite}\ \emph {et~al.}(2022)\citenamefont
  {Attaccalite}, \citenamefont {Prete}, \citenamefont {Palummo},\ and\
  \citenamefont {Pulci}}]{AlN-exciton1}%
  \BibitemOpen
  \bibfield  {author} {\bibinfo {author} {\bibfnamefont {C.}~\bibnamefont
  {Attaccalite}}, \bibinfo {author} {\bibfnamefont {M.~S.}\ \bibnamefont
  {Prete}}, \bibinfo {author} {\bibfnamefont {M.}~\bibnamefont {Palummo}},\
  and\ \bibinfo {author} {\bibfnamefont {O.}~\bibnamefont {Pulci}},\ }\bibfield
   {title} {\bibinfo {title} {Interlayer and intralayer excitons in aln/ws2
  heterostructure},\ }\href {http://dx.doi.org/10.3390/ma15238318} {\bibfield
  {journal} {\bibinfo  {journal} {Materials}\ }\textbf {\bibinfo {volume}
  {15}},\ \bibinfo {pages} {8318} (\bibinfo {year} {2022})}\BibitemShut
  {NoStop}%
\bibitem [{\citenamefont {Helmrich}\ \emph {et~al.}(2018)\citenamefont
  {Helmrich}, \citenamefont {Schneider}, \citenamefont {Achtstein},
  \citenamefont {Arora}, \citenamefont {Herzog}, \citenamefont
  {de~Vasconcellos}, \citenamefont {Kolarczik}, \citenamefont {Schöps},
  \citenamefont {Bratschitsch}, \citenamefont {Woggon},\ and\ \citenamefont
  {Owschimikow}}]{Helmrich_2018}%
  \BibitemOpen
  \bibfield  {author} {\bibinfo {author} {\bibfnamefont {S.}~\bibnamefont
  {Helmrich}}, \bibinfo {author} {\bibfnamefont {R.}~\bibnamefont {Schneider}},
  \bibinfo {author} {\bibfnamefont {A.~W.}\ \bibnamefont {Achtstein}}, \bibinfo
  {author} {\bibfnamefont {A.}~\bibnamefont {Arora}}, \bibinfo {author}
  {\bibfnamefont {B.}~\bibnamefont {Herzog}}, \bibinfo {author} {\bibfnamefont
  {S.~M.}\ \bibnamefont {de~Vasconcellos}}, \bibinfo {author} {\bibfnamefont
  {M.}~\bibnamefont {Kolarczik}}, \bibinfo {author} {\bibfnamefont
  {O.}~\bibnamefont {Schöps}}, \bibinfo {author} {\bibfnamefont
  {R.}~\bibnamefont {Bratschitsch}}, \bibinfo {author} {\bibfnamefont
  {U.}~\bibnamefont {Woggon}},\ and\ \bibinfo {author} {\bibfnamefont
  {N.}~\bibnamefont {Owschimikow}},\ }\bibfield  {title} {\bibinfo {title}
  {Exciton–phonon coupling in mono- and bilayer mote2},\ }\href
  {https://doi.org/10.1088/2053-1583/aacfb7} {\bibfield  {journal} {\bibinfo
  {journal} {2D Materials}\ }\textbf {\bibinfo {volume} {5}},\ \bibinfo {pages}
  {045007} (\bibinfo {year} {2018})}\BibitemShut {NoStop}%
\bibitem [{\citenamefont {Cannuccia}\ \emph {et~al.}(2019)\citenamefont
  {Cannuccia}, \citenamefont {Monserrat},\ and\ \citenamefont
  {Attaccalite}}]{cannuccia2019theory}%
  \BibitemOpen
  \bibfield  {author} {\bibinfo {author} {\bibfnamefont {E.}~\bibnamefont
  {Cannuccia}}, \bibinfo {author} {\bibfnamefont {B.}~\bibnamefont
  {Monserrat}},\ and\ \bibinfo {author} {\bibfnamefont {C.}~\bibnamefont
  {Attaccalite}},\ }\bibfield  {title} {\bibinfo {title} {Theory of
  phonon-assisted luminescence in solids: application to hexagonal boron
  nitride},\ }\href {https://doi.org/10.1103/PhysRevB.99.081109} {\bibfield
  {journal} {\bibinfo  {journal} {Phys. Rev. B}\ }\textbf {\bibinfo {volume}
  {99}},\ \bibinfo {pages} {081109} (\bibinfo {year} {2019})}\BibitemShut
  {NoStop}%
\bibitem [{\citenamefont {Chen}\ \emph {et~al.}(2020)\citenamefont {Chen},
  \citenamefont {Liu}, \citenamefont {Zeng}, \citenamefont {Lu}, \citenamefont
  {Lv}, \citenamefont {Chang}, \citenamefont {Lan}, \citenamefont {Wei},
  \citenamefont {Sun}, \citenamefont {Gao}, \citenamefont {Wang},\ and\
  \citenamefont {Fu}}]{Chen2020}%
  \BibitemOpen
  \bibfield  {author} {\bibinfo {author} {\bibfnamefont {Y.}~\bibnamefont
  {Chen}}, \bibinfo {author} {\bibfnamefont {J.}~\bibnamefont {Liu}}, \bibinfo
  {author} {\bibfnamefont {M.}~\bibnamefont {Zeng}}, \bibinfo {author}
  {\bibfnamefont {F.}~\bibnamefont {Lu}}, \bibinfo {author} {\bibfnamefont
  {T.}~\bibnamefont {Lv}}, \bibinfo {author} {\bibfnamefont {Y.}~\bibnamefont
  {Chang}}, \bibinfo {author} {\bibfnamefont {H.}~\bibnamefont {Lan}}, \bibinfo
  {author} {\bibfnamefont {B.}~\bibnamefont {Wei}}, \bibinfo {author}
  {\bibfnamefont {R.}~\bibnamefont {Sun}}, \bibinfo {author} {\bibfnamefont
  {J.}~\bibnamefont {Gao}}, \bibinfo {author} {\bibfnamefont {Z.}~\bibnamefont
  {Wang}},\ and\ \bibinfo {author} {\bibfnamefont {L.}~\bibnamefont {Fu}},\
  }\bibfield  {title} {\bibinfo {title} {Universal growth of ultra-thin
  {III}{\textendash}v semiconductor single crystals},\ }\href
  {https://doi.org/10.1038/s41467-020-17693-5} {\bibfield  {journal} {\bibinfo
  {journal} {Nature Communications}\ }\textbf {\bibinfo {volume} {11}}
  (\bibinfo {year} {2020})}\BibitemShut {NoStop}%
\bibitem [{\citenamefont {Ferreira}\ \emph {et~al.}(2022)\citenamefont
  {Ferreira}, \citenamefont {Rosati}, \citenamefont {Fitzgerald},\ and\
  \citenamefont {Malic}}]{Ferreira_2023}%
  \BibitemOpen
  \bibfield  {author} {\bibinfo {author} {\bibfnamefont {B.}~\bibnamefont
  {Ferreira}}, \bibinfo {author} {\bibfnamefont {R.}~\bibnamefont {Rosati}},
  \bibinfo {author} {\bibfnamefont {J.~M.}\ \bibnamefont {Fitzgerald}},\ and\
  \bibinfo {author} {\bibfnamefont {E.}~\bibnamefont {Malic}},\ }\bibfield
  {title} {\bibinfo {title} {Signatures of dark excitons in exciton–polariton
  optics of transition metal dichalcogenides},\ }\href
  {https://doi.org/10.1088/2053-1583/aca211} {\bibfield  {journal} {\bibinfo
  {journal} {2D Materials}\ }\textbf {\bibinfo {volume} {10}},\ \bibinfo
  {pages} {015012} (\bibinfo {year} {2022})}\BibitemShut {NoStop}%
\bibitem [{\citenamefont {Park}\ \emph {et~al.}(2018)\citenamefont {Park},
  \citenamefont {Mutz}, \citenamefont {Schultz}, \citenamefont {Blumstengel},
  \citenamefont {Han}, \citenamefont {Aljarb}, \citenamefont {Li},
  \citenamefont {List-Kratochvil}, \citenamefont {Amsalem},\ and\ \citenamefont
  {Koch}}]{Park2018}%
  \BibitemOpen
  \bibfield  {author} {\bibinfo {author} {\bibfnamefont {S.}~\bibnamefont
  {Park}}, \bibinfo {author} {\bibfnamefont {N.}~\bibnamefont {Mutz}}, \bibinfo
  {author} {\bibfnamefont {T.}~\bibnamefont {Schultz}}, \bibinfo {author}
  {\bibfnamefont {S.}~\bibnamefont {Blumstengel}}, \bibinfo {author}
  {\bibfnamefont {A.}~\bibnamefont {Han}}, \bibinfo {author} {\bibfnamefont
  {A.}~\bibnamefont {Aljarb}}, \bibinfo {author} {\bibfnamefont {L.-J.}\
  \bibnamefont {Li}}, \bibinfo {author} {\bibfnamefont {E.~J.~W.}\ \bibnamefont
  {List-Kratochvil}}, \bibinfo {author} {\bibfnamefont {P.}~\bibnamefont
  {Amsalem}},\ and\ \bibinfo {author} {\bibfnamefont {N.}~\bibnamefont
  {Koch}},\ }\bibfield  {title} {\bibinfo {title} {Direct determination of
  monolayer mos$_2$ and wse$_2$ exciton binding energies on insulating and
  metallic substrate},\ }\href {https://doi.org/10.1088/2053-1583/aaa4ca}
  {\bibfield  {journal} {\bibinfo  {journal} {2D Mater.}\ }\textbf {\bibinfo
  {volume} {5}},\ \bibinfo {pages} {025003} (\bibinfo {year}
  {2018})}\BibitemShut {NoStop}%
\bibitem [{\citenamefont {Laxmi}\ \emph {et~al.}(2021)\citenamefont {Laxmi},
  \citenamefont {Dong}, \citenamefont {Wang}, \citenamefont {Qi}, \citenamefont
  {Hao}, \citenamefont {Ouyang}, \citenamefont {Ahmad}, \citenamefont {Shah},
  \citenamefont {Yuan},\ and\ \citenamefont {Zhang}}]{LAXMI2021148089}%
  \BibitemOpen
  \bibfield  {author} {\bibinfo {author} {\bibfnamefont {V.}~\bibnamefont
  {Laxmi}}, \bibinfo {author} {\bibfnamefont {W.}~\bibnamefont {Dong}},
  \bibinfo {author} {\bibfnamefont {H.}~\bibnamefont {Wang}}, \bibinfo {author}
  {\bibfnamefont {D.}~\bibnamefont {Qi}}, \bibinfo {author} {\bibfnamefont
  {Q.}~\bibnamefont {Hao}}, \bibinfo {author} {\bibfnamefont {Z.}~\bibnamefont
  {Ouyang}}, \bibinfo {author} {\bibfnamefont {W.}~\bibnamefont {Ahmad}},
  \bibinfo {author} {\bibfnamefont {M.~N.~U.}\ \bibnamefont {Shah}}, \bibinfo
  {author} {\bibfnamefont {Q.}~\bibnamefont {Yuan}},\ and\ \bibinfo {author}
  {\bibfnamefont {W.}~\bibnamefont {Zhang}},\ }\bibfield  {title} {\bibinfo
  {title} {Protecting black phosphorus with selectively adsorbed graphene
  quantum dot layers},\ }\href
  {https://doi.org/https://doi.org/10.1016/j.apsusc.2020.148089} {\bibfield
  {journal} {\bibinfo  {journal} {Applied Surface Science}\ }\textbf {\bibinfo
  {volume} {538}},\ \bibinfo {pages} {148089} (\bibinfo {year}
  {2021})}\BibitemShut {NoStop}%
\bibitem [{\citenamefont {Allen}\ and\ \citenamefont
  {Heine}(1976)}]{Allen1976}%
  \BibitemOpen
  \bibfield  {author} {\bibinfo {author} {\bibfnamefont {P.~B.}\ \bibnamefont
  {Allen}}\ and\ \bibinfo {author} {\bibfnamefont {V.}~\bibnamefont {Heine}},\
  }\bibfield  {title} {\bibinfo {title} {Theory of the temperature dependence
  of electronic band structures},\ }\href
  {https://doi.org/10.1103/PhysRevB.23.1495} {\bibfield  {journal} {\bibinfo
  {journal} {J. Phys. C}\ }\textbf {\bibinfo {volume} {9}},\ \bibinfo {pages}
  {2305} (\bibinfo {year} {1976})}\BibitemShut {NoStop}%
\bibitem [{\citenamefont {Allen}\ and\ \citenamefont
  {Cardona}(1983)}]{Allen1983}%
  \BibitemOpen
  \bibfield  {author} {\bibinfo {author} {\bibfnamefont {P.~B.}\ \bibnamefont
  {Allen}}\ and\ \bibinfo {author} {\bibfnamefont {M.}~\bibnamefont
  {Cardona}},\ }\bibfield  {title} {\bibinfo {title} {Temperature dependence of
  the direct gap of si and ge},\ }\href
  {https://doi.org/https://doi.org/10.1103/PhysRevB.27.4760} {\bibfield
  {journal} {\bibinfo  {journal} {Phys. Rev. B}\ }\textbf {\bibinfo {volume}
  {27}},\ \bibinfo {pages} {4760} (\bibinfo {year} {1983})}\BibitemShut
  {NoStop}%
\bibitem [{\citenamefont {Poonia}\ \emph {et~al.}(2023)\citenamefont {Poonia},
  \citenamefont {Yadav}, \citenamefont {Mondal}, \citenamefont {Mandal},
  \citenamefont {Taank}, \citenamefont {Shrivastava}, \citenamefont {Nag},
  \citenamefont {Agarwal},\ and\ \citenamefont {Adarsh}}]{Ajay2023}%
  \BibitemOpen
  \bibfield  {author} {\bibinfo {author} {\bibfnamefont {A.~K.}\ \bibnamefont
  {Poonia}}, \bibinfo {author} {\bibfnamefont {P.}~\bibnamefont {Yadav}},
  \bibinfo {author} {\bibfnamefont {B.}~\bibnamefont {Mondal}}, \bibinfo
  {author} {\bibfnamefont {D.}~\bibnamefont {Mandal}}, \bibinfo {author}
  {\bibfnamefont {P.}~\bibnamefont {Taank}}, \bibinfo {author} {\bibfnamefont
  {M.}~\bibnamefont {Shrivastava}}, \bibinfo {author} {\bibfnamefont
  {A.}~\bibnamefont {Nag}}, \bibinfo {author} {\bibfnamefont {A.}~\bibnamefont
  {Agarwal}},\ and\ \bibinfo {author} {\bibfnamefont {K.}~\bibnamefont
  {Adarsh}},\ }\bibfield  {title} {\bibinfo {title} {Room-temperature
  electron-hole condensation in direct-band-gap semiconductor nanocrystals},\
  }\href {https://doi.org/10.1103/PhysRevApplied.20.L021002} {\bibfield
  {journal} {\bibinfo  {journal} {Phys. Rev. Appl.}\ }\textbf {\bibinfo
  {volume} {20}},\ \bibinfo {pages} {L021002} (\bibinfo {year}
  {2023})}\BibitemShut {NoStop}%
\bibitem [{\citenamefont {Lautenschlager}\ \emph {et~al.}(1985)\citenamefont
  {Lautenschlager}, \citenamefont {Allen},\ and\ \citenamefont
  {Cardona}}]{Lautenschlager1985}%
  \BibitemOpen
  \bibfield  {author} {\bibinfo {author} {\bibfnamefont {P.}~\bibnamefont
  {Lautenschlager}}, \bibinfo {author} {\bibfnamefont {P.~B.}\ \bibnamefont
  {Allen}},\ and\ \bibinfo {author} {\bibfnamefont {M.}~\bibnamefont
  {Cardona}},\ }\bibfield  {title} {\bibinfo {title} {Temperature dependence of
  band gaps in si and ge},\ }\href {https://doi.org/10.1103/PhysRevB.31.2163}
  {\bibfield  {journal} {\bibinfo  {journal} {Phys. Rev. B}\ }\textbf {\bibinfo
  {volume} {31}},\ \bibinfo {pages} {2163} (\bibinfo {year}
  {1985})}\BibitemShut {NoStop}%
\bibitem [{\citenamefont {Lautenschlager}\ \emph
  {et~al.}(1987{\natexlab{a}})\citenamefont {Lautenschlager}, \citenamefont
  {Garriga}, \citenamefont {Vina},\ and\ \citenamefont
  {Cardona}}]{Lautenschlager1987}%
  \BibitemOpen
  \bibfield  {author} {\bibinfo {author} {\bibfnamefont {P.}~\bibnamefont
  {Lautenschlager}}, \bibinfo {author} {\bibfnamefont {M.}~\bibnamefont
  {Garriga}}, \bibinfo {author} {\bibfnamefont {L.}~\bibnamefont {Vina}},\ and\
  \bibinfo {author} {\bibfnamefont {M.}~\bibnamefont {Cardona}},\ }\bibfield
  {title} {\bibinfo {title} {Temperature dependence of the dielectric function
  and interband critical points in silicon},\ }\href
  {https://doi.org/10.1103/PhysRevB.36.4821} {\bibfield  {journal} {\bibinfo
  {journal} {Phys. Rev. B}\ }\textbf {\bibinfo {volume} {36}},\ \bibinfo
  {pages} {4821} (\bibinfo {year} {1987}{\natexlab{a}})}\BibitemShut {NoStop}%
\bibitem [{\citenamefont {Zollner}\ \emph {et~al.}(1992)\citenamefont
  {Zollner}, \citenamefont {Cardona},\ and\ \citenamefont
  {Gopalan}}]{Zollner1992}%
  \BibitemOpen
  \bibfield  {author} {\bibinfo {author} {\bibfnamefont {S.}~\bibnamefont
  {Zollner}}, \bibinfo {author} {\bibfnamefont {M.}~\bibnamefont {Cardona}},\
  and\ \bibinfo {author} {\bibfnamefont {S.}~\bibnamefont {Gopalan}},\
  }\bibfield  {title} {\bibinfo {title} {Isotope and temperature shifts of
  direct and indirect band gaps in diamond-type semiconductors},\ }\href
  {https://doi.org/10.1103/PhysRevB.45.3376} {\bibfield  {journal} {\bibinfo
  {journal} {Phys. Rev. B}\ }\textbf {\bibinfo {volume} {45}},\ \bibinfo
  {pages} {3376} (\bibinfo {year} {1992})}\BibitemShut {NoStop}%
\bibitem [{\citenamefont {Lautenschlager}\ \emph
  {et~al.}(1987{\natexlab{b}})\citenamefont {Lautenschlager}, \citenamefont
  {Garriga}, \citenamefont {Logothetidis},\ and\ \citenamefont
  {Cardona}}]{Lautenschlager1987_}%
  \BibitemOpen
  \bibfield  {author} {\bibinfo {author} {\bibfnamefont {P.}~\bibnamefont
  {Lautenschlager}}, \bibinfo {author} {\bibfnamefont {M.}~\bibnamefont
  {Garriga}}, \bibinfo {author} {\bibfnamefont {S.}~\bibnamefont
  {Logothetidis}},\ and\ \bibinfo {author} {\bibfnamefont {M.}~\bibnamefont
  {Cardona}},\ }\bibfield  {title} {\bibinfo {title} {Interband critical points
  of gaas and their temperature dependence},\ }\href
  {https://doi.org/https://doi.org/10.1103/PhysRevB.35.9174} {\bibfield
  {journal} {\bibinfo  {journal} {Phys. Rev. B}\ }\textbf {\bibinfo {volume}
  {35}},\ \bibinfo {pages} {9174} (\bibinfo {year}
  {1987}{\natexlab{b}})}\BibitemShut {NoStop}%
\bibitem [{\citenamefont {Gali}\ \emph {et~al.}(2016)\citenamefont {Gali},
  \citenamefont {Demján}, \citenamefont {V\"{o}r\"{o}s}, \citenamefont
  {Thiering}, \citenamefont {Cannuccia},\ and\ \citenamefont
  {Marini}}]{Gali2016}%
  \BibitemOpen
  \bibfield  {author} {\bibinfo {author} {\bibfnamefont {A.}~\bibnamefont
  {Gali}}, \bibinfo {author} {\bibfnamefont {T.}~\bibnamefont {Demján}},
  \bibinfo {author} {\bibfnamefont {M.}~\bibnamefont {V\"{o}r\"{o}s}}, \bibinfo
  {author} {\bibfnamefont {G.}~\bibnamefont {Thiering}}, \bibinfo {author}
  {\bibfnamefont {E.}~\bibnamefont {Cannuccia}},\ and\ \bibinfo {author}
  {\bibfnamefont {A.}~\bibnamefont {Marini}},\ }\bibfield  {title} {\bibinfo
  {title} {Electron–vibration coupling induced renormalization in the
  photoemission spectrum of diamondoids},\ }\href
  {http://dx.doi.org/10.1038/ncomms11327} {\bibfield  {journal} {\bibinfo
  {journal} {Nature Communications}\ }\textbf {\bibinfo {volume} {7}} (\bibinfo
  {year} {2016})}\BibitemShut {NoStop}%
\bibitem [{\citenamefont {Antonius}\ \emph {et~al.}(2015)\citenamefont
  {Antonius}, \citenamefont {Ponc\'e}, \citenamefont {Lantagne-Hurtubise},
  \citenamefont {Auclair}, \citenamefont {Gonze},\ and\ \citenamefont
  {C\^ot\'e}}]{Antonius2015}%
  \BibitemOpen
  \bibfield  {author} {\bibinfo {author} {\bibfnamefont {G.}~\bibnamefont
  {Antonius}}, \bibinfo {author} {\bibfnamefont {S.}~\bibnamefont {Ponc\'e}},
  \bibinfo {author} {\bibfnamefont {E.}~\bibnamefont {Lantagne-Hurtubise}},
  \bibinfo {author} {\bibfnamefont {G.}~\bibnamefont {Auclair}}, \bibinfo
  {author} {\bibfnamefont {X.}~\bibnamefont {Gonze}},\ and\ \bibinfo {author}
  {\bibfnamefont {M.}~\bibnamefont {C\^ot\'e}},\ }\bibfield  {title} {\bibinfo
  {title} {Dynamical and anharmonic effects on the electron-phonon coupling and
  the zero-point renormalization of the electronic structure},\ }\href
  {https://doi.org/10.1103/PhysRevB.92.085137} {\bibfield  {journal} {\bibinfo
  {journal} {Phys. Rev. B}\ }\textbf {\bibinfo {volume} {92}},\ \bibinfo
  {pages} {085137} (\bibinfo {year} {2015})}\BibitemShut {NoStop}%
\bibitem [{\citenamefont {Trovatello}\ \emph {et~al.}(2020)\citenamefont
  {Trovatello}, \citenamefont {Miranda}, \citenamefont {Molina-Sanchez},
  \citenamefont {Borrego-Varillas}, \citenamefont {Manzoni}, \citenamefont
  {Moretti}, \citenamefont {Ganzer}, \citenamefont {Maiuri}, \citenamefont
  {Wang}, \citenamefont {Dumcenco}, \citenamefont {Kis}, \citenamefont {Wirtz},
  \citenamefont {Marini}, \citenamefont {Soavi}, \citenamefont {Ferrari},
  \citenamefont {Cerullo}, \citenamefont {Sangalli},\ and\ \citenamefont
  {Conte}}]{Chiara2020}%
  \BibitemOpen
  \bibfield  {author} {\bibinfo {author} {\bibfnamefont {C.}~\bibnamefont
  {Trovatello}}, \bibinfo {author} {\bibfnamefont {H.~P.~C.}\ \bibnamefont
  {Miranda}}, \bibinfo {author} {\bibfnamefont {A.}~\bibnamefont
  {Molina-Sanchez}}, \bibinfo {author} {\bibfnamefont {R.}~\bibnamefont
  {Borrego-Varillas}}, \bibinfo {author} {\bibfnamefont {C.}~\bibnamefont
  {Manzoni}}, \bibinfo {author} {\bibfnamefont {L.}~\bibnamefont {Moretti}},
  \bibinfo {author} {\bibfnamefont {L.}~\bibnamefont {Ganzer}}, \bibinfo
  {author} {\bibfnamefont {M.}~\bibnamefont {Maiuri}}, \bibinfo {author}
  {\bibfnamefont {J.}~\bibnamefont {Wang}}, \bibinfo {author} {\bibfnamefont
  {D.}~\bibnamefont {Dumcenco}}, \bibinfo {author} {\bibfnamefont
  {A.}~\bibnamefont {Kis}}, \bibinfo {author} {\bibfnamefont {L.}~\bibnamefont
  {Wirtz}}, \bibinfo {author} {\bibfnamefont {A.}~\bibnamefont {Marini}},
  \bibinfo {author} {\bibfnamefont {G.}~\bibnamefont {Soavi}}, \bibinfo
  {author} {\bibfnamefont {A.~C.}\ \bibnamefont {Ferrari}}, \bibinfo {author}
  {\bibfnamefont {G.}~\bibnamefont {Cerullo}}, \bibinfo {author} {\bibfnamefont
  {D.}~\bibnamefont {Sangalli}},\ and\ \bibinfo {author} {\bibfnamefont
  {S.~D.}\ \bibnamefont {Conte}},\ }\bibfield  {title} {\bibinfo {title}
  {Strongly coupled coherent phonons in single-layer mos2},\ }\href
  {https://doi.org/10.1021/acsnano.0c00309} {\bibfield  {journal} {\bibinfo
  {journal} {{ACS} Nano}\ }\textbf {\bibinfo {volume} {14}},\ \bibinfo {pages}
  {5700} (\bibinfo {year} {2020})}\BibitemShut {NoStop}%
\bibitem [{\citenamefont {hao Chan}\ \emph {et~al.}(2023)\citenamefont {hao
  Chan}, \citenamefont {Haber}, \citenamefont {Naik}, \citenamefont {Neaton},
  \citenamefont {Qiu}, \citenamefont {da~Jornada},\ and\ \citenamefont
  {Louie}}]{Chan23}%
  \BibitemOpen
  \bibfield  {author} {\bibinfo {author} {\bibfnamefont {Y.}~\bibnamefont {hao
  Chan}}, \bibinfo {author} {\bibfnamefont {J.~B.}\ \bibnamefont {Haber}},
  \bibinfo {author} {\bibfnamefont {M.~H.}\ \bibnamefont {Naik}}, \bibinfo
  {author} {\bibfnamefont {J.~B.}\ \bibnamefont {Neaton}}, \bibinfo {author}
  {\bibfnamefont {D.~Y.}\ \bibnamefont {Qiu}}, \bibinfo {author} {\bibfnamefont
  {F.~H.}\ \bibnamefont {da~Jornada}},\ and\ \bibinfo {author} {\bibfnamefont
  {S.~G.}\ \bibnamefont {Louie}},\ }\bibfield  {title} {\bibinfo {title}
  {Exciton lifetime and optical line width profile via exciton-phonon
  interactions: Theory and first-principles calculations for monolayer
  {MoS}$_2$},\ }\href {https://doi.org/10.1021/acs.nanolett.3c00732} {\bibfield
   {journal} {\bibinfo  {journal} {Nano Letters}\ }\textbf {\bibinfo {volume}
  {23}},\ \bibinfo {pages} {3971} (\bibinfo {year} {2023})}\BibitemShut
  {NoStop}%
\bibitem [{\citenamefont {Mishra}\ \emph {et~al.}(2018)\citenamefont {Mishra},
  \citenamefont {Bose}, \citenamefont {Dhar},\ and\ \citenamefont
  {Bhattacharya}}]{HimaniWSe2}%
  \BibitemOpen
  \bibfield  {author} {\bibinfo {author} {\bibfnamefont {H.}~\bibnamefont
  {Mishra}}, \bibinfo {author} {\bibfnamefont {A.}~\bibnamefont {Bose}},
  \bibinfo {author} {\bibfnamefont {A.}~\bibnamefont {Dhar}},\ and\ \bibinfo
  {author} {\bibfnamefont {S.}~\bibnamefont {Bhattacharya}},\ }\bibfield
  {title} {\bibinfo {title} {Exciton-phonon coupling and band-gap
  renormalization in monolayer ${\mathrm{wse}}_{2}$},\ }\href
  {https://doi.org/10.1103/PhysRevB.98.045143} {\bibfield  {journal} {\bibinfo
  {journal} {Phys. Rev. B}\ }\textbf {\bibinfo {volume} {98}},\ \bibinfo
  {pages} {045143} (\bibinfo {year} {2018})}\BibitemShut {NoStop}%
\bibitem [{\citenamefont {Mishra}\ and\ \citenamefont
  {Bhattacharya}(2019)}]{Mishra2019}%
  \BibitemOpen
  \bibfield  {author} {\bibinfo {author} {\bibfnamefont {H.}~\bibnamefont
  {Mishra}}\ and\ \bibinfo {author} {\bibfnamefont {S.}~\bibnamefont
  {Bhattacharya}},\ }\bibfield  {title} {\bibinfo {title} {Giant exciton-phonon
  coupling and zero-point renormalization in hexagonal monolayer boron
  nitride},\ }\href
  {https://doi.org/https://doi.org/10.1103/PhysRevB.99.165201} {\bibfield
  {journal} {\bibinfo  {journal} {Phys. Rev. B}\ }\textbf {\bibinfo {volume}
  {99}},\ \bibinfo {pages} {165201} (\bibinfo {year} {2019})}\BibitemShut
  {NoStop}%
\bibitem [{\citenamefont {Kolos}\ \emph {et~al.}(2021)\citenamefont {Kolos},
  \citenamefont {Cigarini}, \citenamefont {Verma}, \citenamefont {Karlický},\
  and\ \citenamefont {Bhattacharya}}]{Kolos2021}%
  \BibitemOpen
  \bibfield  {author} {\bibinfo {author} {\bibfnamefont {M.}~\bibnamefont
  {Kolos}}, \bibinfo {author} {\bibfnamefont {L.}~\bibnamefont {Cigarini}},
  \bibinfo {author} {\bibfnamefont {R.}~\bibnamefont {Verma}}, \bibinfo
  {author} {\bibfnamefont {F.}~\bibnamefont {Karlický}},\ and\ \bibinfo
  {author} {\bibfnamefont {S.}~\bibnamefont {Bhattacharya}},\ }\bibfield
  {title} {\bibinfo {title} {Giant linear and nonlinear excitonic responses in
  an atomically thin indirect semiconductor nitrogen phosphide},\ }\href
  {https://doi.org/10.1021/acs.jpcc.1c02091} {\bibfield  {journal} {\bibinfo
  {journal} {The Journal of Physical Chemistry C}\ }\textbf {\bibinfo {volume}
  {125}},\ \bibinfo {pages} {12738} (\bibinfo {year} {2021})}\BibitemShut
  {NoStop}%
\bibitem [{\citenamefont {Zhuang}\ \emph
  {et~al.}(2013{\natexlab{b}})\citenamefont {Zhuang}, \citenamefont {Singh},\
  and\ \citenamefont {Hennig}}]{ZhuangAlN}%
  \BibitemOpen
  \bibfield  {author} {\bibinfo {author} {\bibfnamefont {H.~L.}\ \bibnamefont
  {Zhuang}}, \bibinfo {author} {\bibfnamefont {A.~K.}\ \bibnamefont {Singh}},\
  and\ \bibinfo {author} {\bibfnamefont {R.~G.}\ \bibnamefont {Hennig}},\
  }\bibfield  {title} {\bibinfo {title} {Computational discovery of
  single-layer iii-v materials},\ }\href
  {https://doi.org/10.1103/PhysRevB.87.165415} {\bibfield  {journal} {\bibinfo
  {journal} {Phys. Rev. B}\ }\textbf {\bibinfo {volume} {87}},\ \bibinfo
  {pages} {165415} (\bibinfo {year} {2013}{\natexlab{b}})}\BibitemShut
  {NoStop}%
\bibitem [{\citenamefont {Giannozzi}\ \emph {et~al.}(2017)\citenamefont
  {Giannozzi}, \citenamefont {Andreussi}, \citenamefont {Brumme}, \citenamefont
  {Bunau}, \citenamefont {Nardelli}, \citenamefont {Calandra}, \citenamefont
  {Car}, \citenamefont {Cavazzoni}, \citenamefont {Ceresoli}, \citenamefont
  {Cococcioni}, \citenamefont {Colonna}, \citenamefont {Carnimeo},
  \citenamefont {Corso}, \citenamefont {de~Gironcoli}, \citenamefont {Delugas},
  \citenamefont {DiStasio}, \citenamefont {Ferretti}, \citenamefont {Floris},
  \citenamefont {Fratesi}, \citenamefont {Fugallo}, \citenamefont {Gebauer},
  \citenamefont {Gerstmann}, \citenamefont {Giustino}, \citenamefont {Gorni},
  \citenamefont {Jia}, \citenamefont {Kawamura}, \citenamefont {Ko},
  \citenamefont {Kokalj}, \citenamefont {Kü{\c{c}}ükbenli}, \citenamefont
  {Lazzeri}, \citenamefont {Marsili}, \citenamefont {Marzari}, \citenamefont
  {Mauri}, \citenamefont {Nguyen}, \citenamefont {Nguyen}, \citenamefont {de-la
  Roza}, \citenamefont {Paulatto}, \citenamefont {Ponc{\'{e}}}, \citenamefont
  {Rocca}, \citenamefont {Sabatini}, \citenamefont {Santra}, \citenamefont
  {Schlipf}, \citenamefont {Seitsonen}, \citenamefont {Smogunov}, \citenamefont
  {Timrov}, \citenamefont {Thonhauser}, \citenamefont {Umari}, \citenamefont
  {Vast}, \citenamefont {Wu},\ and\ \citenamefont {Baroni}}]{QE}%
  \BibitemOpen
  \bibfield  {author} {\bibinfo {author} {\bibfnamefont {P.}~\bibnamefont
  {Giannozzi}}, \bibinfo {author} {\bibfnamefont {O.}~\bibnamefont
  {Andreussi}}, \bibinfo {author} {\bibfnamefont {T.}~\bibnamefont {Brumme}},
  \bibinfo {author} {\bibfnamefont {O.}~\bibnamefont {Bunau}}, \bibinfo
  {author} {\bibfnamefont {M.~B.}\ \bibnamefont {Nardelli}}, \bibinfo {author}
  {\bibfnamefont {M.}~\bibnamefont {Calandra}}, \bibinfo {author}
  {\bibfnamefont {R.}~\bibnamefont {Car}}, \bibinfo {author} {\bibfnamefont
  {C.}~\bibnamefont {Cavazzoni}}, \bibinfo {author} {\bibfnamefont
  {D.}~\bibnamefont {Ceresoli}}, \bibinfo {author} {\bibfnamefont
  {M.}~\bibnamefont {Cococcioni}}, \bibinfo {author} {\bibfnamefont
  {N.}~\bibnamefont {Colonna}}, \bibinfo {author} {\bibfnamefont
  {I.}~\bibnamefont {Carnimeo}}, \bibinfo {author} {\bibfnamefont {A.~D.}\
  \bibnamefont {Corso}}, \bibinfo {author} {\bibfnamefont {S.}~\bibnamefont
  {de~Gironcoli}}, \bibinfo {author} {\bibfnamefont {P.}~\bibnamefont
  {Delugas}}, \bibinfo {author} {\bibfnamefont {R.~A.}\ \bibnamefont
  {DiStasio}}, \bibinfo {author} {\bibfnamefont {A.}~\bibnamefont {Ferretti}},
  \bibinfo {author} {\bibfnamefont {A.}~\bibnamefont {Floris}}, \bibinfo
  {author} {\bibfnamefont {G.}~\bibnamefont {Fratesi}}, \bibinfo {author}
  {\bibfnamefont {G.}~\bibnamefont {Fugallo}}, \bibinfo {author} {\bibfnamefont
  {R.}~\bibnamefont {Gebauer}}, \bibinfo {author} {\bibfnamefont
  {U.}~\bibnamefont {Gerstmann}}, \bibinfo {author} {\bibfnamefont
  {F.}~\bibnamefont {Giustino}}, \bibinfo {author} {\bibfnamefont
  {T.}~\bibnamefont {Gorni}}, \bibinfo {author} {\bibfnamefont
  {J.}~\bibnamefont {Jia}}, \bibinfo {author} {\bibfnamefont {M.}~\bibnamefont
  {Kawamura}}, \bibinfo {author} {\bibfnamefont {H.-Y.}\ \bibnamefont {Ko}},
  \bibinfo {author} {\bibfnamefont {A.}~\bibnamefont {Kokalj}}, \bibinfo
  {author} {\bibfnamefont {E.}~\bibnamefont {Kü{\c{c}}ükbenli}}, \bibinfo
  {author} {\bibfnamefont {M.}~\bibnamefont {Lazzeri}}, \bibinfo {author}
  {\bibfnamefont {M.}~\bibnamefont {Marsili}}, \bibinfo {author} {\bibfnamefont
  {N.}~\bibnamefont {Marzari}}, \bibinfo {author} {\bibfnamefont
  {F.}~\bibnamefont {Mauri}}, \bibinfo {author} {\bibfnamefont {N.~L.}\
  \bibnamefont {Nguyen}}, \bibinfo {author} {\bibfnamefont {H.-V.}\
  \bibnamefont {Nguyen}}, \bibinfo {author} {\bibfnamefont {A.~O.}\
  \bibnamefont {de-la Roza}}, \bibinfo {author} {\bibfnamefont
  {L.}~\bibnamefont {Paulatto}}, \bibinfo {author} {\bibfnamefont
  {S.}~\bibnamefont {Ponc{\'{e}}}}, \bibinfo {author} {\bibfnamefont
  {D.}~\bibnamefont {Rocca}}, \bibinfo {author} {\bibfnamefont
  {R.}~\bibnamefont {Sabatini}}, \bibinfo {author} {\bibfnamefont
  {B.}~\bibnamefont {Santra}}, \bibinfo {author} {\bibfnamefont
  {M.}~\bibnamefont {Schlipf}}, \bibinfo {author} {\bibfnamefont {A.~P.}\
  \bibnamefont {Seitsonen}}, \bibinfo {author} {\bibfnamefont {A.}~\bibnamefont
  {Smogunov}}, \bibinfo {author} {\bibfnamefont {I.}~\bibnamefont {Timrov}},
  \bibinfo {author} {\bibfnamefont {T.}~\bibnamefont {Thonhauser}}, \bibinfo
  {author} {\bibfnamefont {P.}~\bibnamefont {Umari}}, \bibinfo {author}
  {\bibfnamefont {N.}~\bibnamefont {Vast}}, \bibinfo {author} {\bibfnamefont
  {X.}~\bibnamefont {Wu}},\ and\ \bibinfo {author} {\bibfnamefont
  {S.}~\bibnamefont {Baroni}},\ }\bibfield  {title} {\bibinfo {title} {Advanced
  capabilities for materials modelling with quantum {ESPRESSO}},\ }\href
  {https://doi.org/10.1088/1361-648x/aa8f79} {\bibfield  {journal} {\bibinfo
  {journal} {Journal of Physics: Condensed Matter}\ }\textbf {\bibinfo {volume}
  {29}},\ \bibinfo {pages} {465901} (\bibinfo {year} {2017})}\BibitemShut
  {NoStop}%
\bibitem [{\citenamefont {Perdew}\ \emph {et~al.}(1996)\citenamefont {Perdew},
  \citenamefont {Burke},\ and\ \citenamefont {Ernzerhof}}]{GGA-PBE}%
  \BibitemOpen
  \bibfield  {author} {\bibinfo {author} {\bibfnamefont {J.~P.}\ \bibnamefont
  {Perdew}}, \bibinfo {author} {\bibfnamefont {K.}~\bibnamefont {Burke}},\ and\
  \bibinfo {author} {\bibfnamefont {M.}~\bibnamefont {Ernzerhof}},\ }\bibfield
  {title} {\bibinfo {title} {Generalized gradient approximation made simple},\
  }\href {https://doi.org/10.1103/PhysRevLett.77.3865} {\bibfield  {journal}
  {\bibinfo  {journal} {Phys. Rev. Lett.}\ }\textbf {\bibinfo {volume} {77}},\
  \bibinfo {pages} {3865} (\bibinfo {year} {1996})}\BibitemShut {NoStop}%
\bibitem [{Sup()}]{Supplemental}%
  \BibitemOpen
  \href@noop {} {}\bibinfo {howpublished} {See Supplemental section at
  http://-/supplemental/ for necessary convergence and supporting
  figures}\BibitemShut {NoStop}%
\bibitem [{\citenamefont {Fan}(1950)}]{Fan1950}%
  \BibitemOpen
  \bibfield  {author} {\bibinfo {author} {\bibfnamefont {H.~Y.}\ \bibnamefont
  {Fan}},\ }\bibfield  {title} {\bibinfo {title} {Temperature dependence of the
  energy gap in monatomic semiconductors},\ }\href
  {https://doi.org/10.1103/PhysRev.78.808.2} {\bibfield  {journal} {\bibinfo
  {journal} {Phys. Rev.}\ }\textbf {\bibinfo {volume} {6}},\ \bibinfo {pages}
  {808} (\bibinfo {year} {1950})}\BibitemShut {NoStop}%
\bibitem [{\citenamefont {Kim}\ \emph {et~al.}(1986)\citenamefont {Kim},
  \citenamefont {Lautenschlager},\ and\ \citenamefont {Cardona}}]{Kim1986}%
  \BibitemOpen
  \bibfield  {author} {\bibinfo {author} {\bibfnamefont {C.~K.}\ \bibnamefont
  {Kim}}, \bibinfo {author} {\bibfnamefont {P.}~\bibnamefont
  {Lautenschlager}},\ and\ \bibinfo {author} {\bibfnamefont {M.}~\bibnamefont
  {Cardona}},\ }\bibfield  {title} {\bibinfo {title} {Temperature dependence of
  the fundamental energy gap in gaas},\ }\href
  {https://doi.org/https://doi.org/10.1016/0038-1098(86)90632-0} {\bibfield
  {journal} {\bibinfo  {journal} {Solid State Commun.}\ }\textbf {\bibinfo
  {volume} {59}},\ \bibinfo {pages} {797} (\bibinfo {year} {1986})}\BibitemShut
  {NoStop}%
\bibitem [{\citenamefont {Mahan}(2000)}]{Mahan2014}%
  \BibitemOpen
  \bibfield  {author} {\bibinfo {author} {\bibfnamefont {G.~D.}\ \bibnamefont
  {Mahan}},\ }\href {https://doi.org/10.1007/978-1-4757-5714-9} {\emph
  {\bibinfo {title} {Many-Particle Physics}}}\ (\bibinfo  {publisher} {Springer
  US},\ \bibinfo {year} {2000})\BibitemShut {NoStop}%
\bibitem [{\citenamefont {Cannuccia}(2011)}]{cannuccia2011}%
  \BibitemOpen
  \bibfield  {author} {\bibinfo {author} {\bibfnamefont {E.}~\bibnamefont
  {Cannuccia}},\ }\href@noop {} {\emph {\bibinfo {title} {Giant polaronic
  effects in polymers: breakdown of the quasiparticle picture}}}\ (\bibinfo
  {publisher} {PhD thesis},\ \bibinfo {address} {Rome Tor Vergata University},\
  \bibinfo {year} {2011})\BibitemShut {NoStop}%
\bibitem [{\citenamefont {Strinati}(1982)}]{BSE-1-Strinati}%
  \BibitemOpen
  \bibfield  {author} {\bibinfo {author} {\bibfnamefont {G.}~\bibnamefont
  {Strinati}},\ }\bibfield  {title} {\bibinfo {title} {Dynamical shift and
  broadening of core excitons in semiconductors},\ }\href
  {https://doi.org/10.1103/PhysRevLett.49.1519} {\bibfield  {journal} {\bibinfo
   {journal} {Phys. Rev. Lett.}\ }\textbf {\bibinfo {volume} {49}},\ \bibinfo
  {pages} {1519} (\bibinfo {year} {1982})}\BibitemShut {NoStop}%
\bibitem [{\citenamefont {Rohlfing}\ and\ \citenamefont
  {Louie}(2005)}]{Rohlfing2000}%
  \BibitemOpen
  \bibfield  {author} {\bibinfo {author} {\bibfnamefont {M.}~\bibnamefont
  {Rohlfing}}\ and\ \bibinfo {author} {\bibfnamefont {S.~G.}\ \bibnamefont
  {Louie}},\ }\bibfield  {title} {\bibinfo {title} {Electron-hole excitations
  and optical spectra from first principles},\ }\href
  {https://doi.org/10.1103/PhysRevB.62.4927} {\bibfield  {journal} {\bibinfo
  {journal} {Phys. Rev. B}\ }\textbf {\bibinfo {volume} {62}},\ \bibinfo
  {pages} {4927} (\bibinfo {year} {2005})}\BibitemShut {NoStop}%
\bibitem [{\citenamefont {Sangalli}\ \emph {et~al.}(2019)\citenamefont
  {Sangalli}, \citenamefont {Ferretti}, \citenamefont {Miranda}, \citenamefont
  {Attaccalite}, \citenamefont {Marri}, \citenamefont {Cannuccia},
  \citenamefont {Melo}, \citenamefont {Marsili}, \citenamefont {Paleari},
  \citenamefont {Marrazzo}, \citenamefont {Prandini}, \citenamefont
  {Bonf{\`{a}}}, \citenamefont {Atambo}, \citenamefont {Affinito},
  \citenamefont {Palummo}, \citenamefont {Molina-S{\'{a}}nchez}, \citenamefont
  {Hogan}, \citenamefont {Grüning}, \citenamefont {Varsano},\ and\
  \citenamefont {Marini}}]{yambo2019}%
  \BibitemOpen
  \bibfield  {author} {\bibinfo {author} {\bibfnamefont {D.}~\bibnamefont
  {Sangalli}}, \bibinfo {author} {\bibfnamefont {A.}~\bibnamefont {Ferretti}},
  \bibinfo {author} {\bibfnamefont {H.}~\bibnamefont {Miranda}}, \bibinfo
  {author} {\bibfnamefont {C.}~\bibnamefont {Attaccalite}}, \bibinfo {author}
  {\bibfnamefont {I.}~\bibnamefont {Marri}}, \bibinfo {author} {\bibfnamefont
  {E.}~\bibnamefont {Cannuccia}}, \bibinfo {author} {\bibfnamefont
  {P.}~\bibnamefont {Melo}}, \bibinfo {author} {\bibfnamefont {M.}~\bibnamefont
  {Marsili}}, \bibinfo {author} {\bibfnamefont {F.}~\bibnamefont {Paleari}},
  \bibinfo {author} {\bibfnamefont {A.}~\bibnamefont {Marrazzo}}, \bibinfo
  {author} {\bibfnamefont {G.}~\bibnamefont {Prandini}}, \bibinfo {author}
  {\bibfnamefont {P.}~\bibnamefont {Bonf{\`{a}}}}, \bibinfo {author}
  {\bibfnamefont {M.~O.}\ \bibnamefont {Atambo}}, \bibinfo {author}
  {\bibfnamefont {F.}~\bibnamefont {Affinito}}, \bibinfo {author}
  {\bibfnamefont {M.}~\bibnamefont {Palummo}}, \bibinfo {author} {\bibfnamefont
  {A.}~\bibnamefont {Molina-S{\'{a}}nchez}}, \bibinfo {author} {\bibfnamefont
  {C.}~\bibnamefont {Hogan}}, \bibinfo {author} {\bibfnamefont
  {M.}~\bibnamefont {Grüning}}, \bibinfo {author} {\bibfnamefont
  {D.}~\bibnamefont {Varsano}},\ and\ \bibinfo {author} {\bibfnamefont
  {A.}~\bibnamefont {Marini}},\ }\bibfield  {title} {\bibinfo {title}
  {Many-body perturbation theory calculations using the yambo code},\ }\href
  {https://doi.org/10.1088/1361-648x/ab15d0} {\bibfield  {journal} {\bibinfo
  {journal} {Journal of Physics: Condensed Matter}\ }\textbf {\bibinfo {volume}
  {31}},\ \bibinfo {pages} {325902} (\bibinfo {year} {2019})}\BibitemShut
  {NoStop}%
\bibitem [{\citenamefont {Aryasetiawan}\ and\ \citenamefont
  {Gunnarsson}(1998)}]{Aryasetiawan1998}%
  \BibitemOpen
  \bibfield  {author} {\bibinfo {author} {\bibfnamefont {F.}~\bibnamefont
  {Aryasetiawan}}\ and\ \bibinfo {author} {\bibfnamefont {O.}~\bibnamefont
  {Gunnarsson}},\ }\bibfield  {title} {\bibinfo {title} {The gw method},\
  }\href {https://doi.org/http://dx.doi.org/10.1088/0034-4885/61/3/002}
  {\bibfield  {journal} {\bibinfo  {journal} {Rep. Prog. Phys.}\ }\textbf
  {\bibinfo {volume} {61}},\ \bibinfo {pages} {237} (\bibinfo {year}
  {1998})}\BibitemShut {NoStop}%
\bibitem [{\citenamefont {Paleari}\ \emph {et~al.}(2018)\citenamefont
  {Paleari}, \citenamefont {Galvani}, \citenamefont {Amara}, \citenamefont
  {Du\c{c}astelle}, \citenamefont {Molina-S{\'a}nchez},\ and\ \citenamefont
  {Wirtz}}]{Paleari2018}%
  \BibitemOpen
  \bibfield  {author} {\bibinfo {author} {\bibfnamefont {F.}~\bibnamefont
  {Paleari}}, \bibinfo {author} {\bibfnamefont {T.}~\bibnamefont {Galvani}},
  \bibinfo {author} {\bibfnamefont {H.}~\bibnamefont {Amara}}, \bibinfo
  {author} {\bibfnamefont {F.}~\bibnamefont {Du\c{c}astelle}}, \bibinfo
  {author} {\bibfnamefont {A.}~\bibnamefont {Molina-S{\'a}nchez}},\ and\
  \bibinfo {author} {\bibfnamefont {L.}~\bibnamefont {Wirtz}},\ }\bibfield
  {title} {\bibinfo {title} {Excitons in few-layer hexagonal boron nitride:
  Davydov splitting and surface localization},\ }\href
  {https://doi.org/10.1088/2053-1583/aad586} {\bibfield  {journal} {\bibinfo
  {journal} {2D Mater.}\ }\textbf {\bibinfo {volume} {5}},\ \bibinfo {pages}
  {045017} (\bibinfo {year} {2018})}\BibitemShut {NoStop}%
\bibitem [{\citenamefont {Miranda}\ \emph {et~al.}(2017)\citenamefont
  {Miranda}, \citenamefont {Reichardt}, \citenamefont {Froehlicher},
  \citenamefont {Molina-S{\'a}nchez}, \citenamefont {Berciaud},\ and\
  \citenamefont {Wirtz}}]{Henrique2017}%
  \BibitemOpen
  \bibfield  {author} {\bibinfo {author} {\bibfnamefont {H.~P.~C.}\
  \bibnamefont {Miranda}}, \bibinfo {author} {\bibfnamefont {S.}~\bibnamefont
  {Reichardt}}, \bibinfo {author} {\bibfnamefont {G.}~\bibnamefont
  {Froehlicher}}, \bibinfo {author} {\bibfnamefont {A.}~\bibnamefont
  {Molina-S{\'a}nchez}}, \bibinfo {author} {\bibfnamefont {S.}~\bibnamefont
  {Berciaud}},\ and\ \bibinfo {author} {\bibfnamefont {L.}~\bibnamefont
  {Wirtz}},\ }\bibfield  {title} {\bibinfo {title} {Excitons in few-layer
  hexagonal boron nitride: Davydov splitting and surface localization},\ }\href
  {https://doi.org/10.1088/2053-1583/aad586} {\bibfield  {journal} {\bibinfo
  {journal} {Nano Lett.}\ }\textbf {\bibinfo {volume} {17}},\ \bibinfo {pages}
  {2381} (\bibinfo {year} {2017})}\BibitemShut {NoStop}%
\bibitem [{\citenamefont {Molina-S\'anchez}\ \emph {et~al.}(2016)\citenamefont
  {Molina-S\'anchez}, \citenamefont {Palummo}, \citenamefont {Marini},\ and\
  \citenamefont {Wirtz}}]{Molina2016}%
  \BibitemOpen
  \bibfield  {author} {\bibinfo {author} {\bibfnamefont {A.}~\bibnamefont
  {Molina-S\'anchez}}, \bibinfo {author} {\bibfnamefont {M.}~\bibnamefont
  {Palummo}}, \bibinfo {author} {\bibfnamefont {A.}~\bibnamefont {Marini}},\
  and\ \bibinfo {author} {\bibfnamefont {L.}~\bibnamefont {Wirtz}},\ }\bibfield
   {title} {\bibinfo {title} {Temperature-dependent excitonic effects in the
  optical properties of single-layer ${\mathrm{mos}}_{2}$},\ }\href
  {https://doi.org/10.1103/PhysRevB.93.155435} {\bibfield  {journal} {\bibinfo
  {journal} {Phys. Rev. B}\ }\textbf {\bibinfo {volume} {93}},\ \bibinfo
  {pages} {155435} (\bibinfo {year} {2016})}\BibitemShut {NoStop}%
\bibitem [{\citenamefont {Purz}\ \emph {et~al.}(2022)\citenamefont {Purz},
  \citenamefont {Martin}, \citenamefont {Holtzmann}, \citenamefont {Rivera},
  \citenamefont {Alfrey}, \citenamefont {Bates}, \citenamefont {Deng},
  \citenamefont {Xu},\ and\ \citenamefont {Cundiff}}]{Purz2022}%
  \BibitemOpen
  \bibfield  {author} {\bibinfo {author} {\bibfnamefont {T.~L.}\ \bibnamefont
  {Purz}}, \bibinfo {author} {\bibfnamefont {E.~W.}\ \bibnamefont {Martin}},
  \bibinfo {author} {\bibfnamefont {W.~G.}\ \bibnamefont {Holtzmann}}, \bibinfo
  {author} {\bibfnamefont {P.}~\bibnamefont {Rivera}}, \bibinfo {author}
  {\bibfnamefont {A.}~\bibnamefont {Alfrey}}, \bibinfo {author} {\bibfnamefont
  {K.~M.}\ \bibnamefont {Bates}}, \bibinfo {author} {\bibfnamefont
  {H.}~\bibnamefont {Deng}}, \bibinfo {author} {\bibfnamefont {X.}~\bibnamefont
  {Xu}},\ and\ \bibinfo {author} {\bibfnamefont {S.~T.}\ \bibnamefont
  {Cundiff}},\ }\bibfield  {title} {\bibinfo {title} {{Imaging dynamic exciton
  interactions and coupling in transition metal dichalcogenides}},\ }\href
  {https://doi.org/10.1063/5.0087544} {\bibfield  {journal} {\bibinfo
  {journal} {The Journal of Chemical Physics}\ }\textbf {\bibinfo {volume}
  {156}},\ \bibinfo {pages} {214704} (\bibinfo {year} {2022})}\BibitemShut
  {NoStop}%
\bibitem [{\citenamefont {Moody}\ \emph {et~al.}(2015)\citenamefont {Moody},
  \citenamefont {Dass}, \citenamefont {Hao}, \citenamefont {Chen},
  \citenamefont {Li}, \citenamefont {Singh}, \citenamefont {Tran},
  \citenamefont {Clark}, \citenamefont {Xu},\ and\ \citenamefont {\textit{et
  al.}}}]{Moody2015}%
  \BibitemOpen
  \bibfield  {author} {\bibinfo {author} {\bibfnamefont {G.}~\bibnamefont
  {Moody}}, \bibinfo {author} {\bibfnamefont {C.~K.}\ \bibnamefont {Dass}},
  \bibinfo {author} {\bibfnamefont {K.}~\bibnamefont {Hao}}, \bibinfo {author}
  {\bibfnamefont {C.-H.}\ \bibnamefont {Chen}}, \bibinfo {author}
  {\bibfnamefont {L.-J.}\ \bibnamefont {Li}}, \bibinfo {author} {\bibfnamefont
  {A.}~\bibnamefont {Singh}}, \bibinfo {author} {\bibfnamefont
  {K.}~\bibnamefont {Tran}}, \bibinfo {author} {\bibfnamefont {G.}~\bibnamefont
  {Clark}}, \bibinfo {author} {\bibfnamefont {X.}~\bibnamefont {Xu}},\ and\
  \bibinfo {author} {\bibfnamefont {G.~B.}\ \bibnamefont {\textit{et al.}}},\
  }\bibfield  {title} {\bibinfo {title} {Intrinsic homogeneous linewidth and
  broadening mechanisms of excitons in monolayer transition metal
  dichalcogenides},\ }\href
  {https://doi.org/https://doi.org/10.1038/ncomms13279} {\bibfield  {journal}
  {\bibinfo  {journal} {Nat. Commun.}\ }\textbf {\bibinfo {volume} {6}},\
  \bibinfo {pages} {8315} (\bibinfo {year} {2015})}\BibitemShut {NoStop}%
\bibitem [{\citenamefont {Dey}\ \emph {et~al.}(2016)\citenamefont {Dey},
  \citenamefont {Paul}, \citenamefont {Wang}, \citenamefont {Stevens},
  \citenamefont {Liu}, \citenamefont {Romero}, \citenamefont {Shan},
  \citenamefont {Hilton},\ and\ \citenamefont {Karaiskaj}}]{dey2016optical}%
  \BibitemOpen
  \bibfield  {author} {\bibinfo {author} {\bibfnamefont {P.}~\bibnamefont
  {Dey}}, \bibinfo {author} {\bibfnamefont {J.}~\bibnamefont {Paul}}, \bibinfo
  {author} {\bibfnamefont {Z.}~\bibnamefont {Wang}}, \bibinfo {author}
  {\bibfnamefont {C.}~\bibnamefont {Stevens}}, \bibinfo {author} {\bibfnamefont
  {C.}~\bibnamefont {Liu}}, \bibinfo {author} {\bibfnamefont {A.}~\bibnamefont
  {Romero}}, \bibinfo {author} {\bibfnamefont {J.}~\bibnamefont {Shan}},
  \bibinfo {author} {\bibfnamefont {D.}~\bibnamefont {Hilton}},\ and\ \bibinfo
  {author} {\bibfnamefont {D.}~\bibnamefont {Karaiskaj}},\ }\bibfield  {title}
  {\bibinfo {title} {Optical coherence in atomic-monolayer transition-metal
  dichalcogenides limited by electron-phonon interactions},\ }\href
  {https://doi.org/10.1103/PhysRevLett.116.127402} {\bibfield  {journal}
  {\bibinfo  {journal} {Physical review letters}\ }\textbf {\bibinfo {volume}
  {116}},\ \bibinfo {pages} {127402} (\bibinfo {year} {2016})}\BibitemShut
  {NoStop}%
\bibitem [{\citenamefont {Selig}\ \emph {et~al.}(2016)\citenamefont {Selig},
  \citenamefont {Bergh{\"a}user}, \citenamefont {Raja}, \citenamefont {Nagler},
  \citenamefont {Sch{\"u}ller}, \citenamefont {Heinz}, \citenamefont {Korn},
  \citenamefont {Chernikov}, \citenamefont {Malic},\ and\ \citenamefont
  {Knorr}}]{Selig2016}%
  \BibitemOpen
  \bibfield  {author} {\bibinfo {author} {\bibfnamefont {M.}~\bibnamefont
  {Selig}}, \bibinfo {author} {\bibfnamefont {G.}~\bibnamefont
  {Bergh{\"a}user}}, \bibinfo {author} {\bibfnamefont {A.}~\bibnamefont
  {Raja}}, \bibinfo {author} {\bibfnamefont {P.}~\bibnamefont {Nagler}},
  \bibinfo {author} {\bibfnamefont {C.}~\bibnamefont {Sch{\"u}ller}}, \bibinfo
  {author} {\bibfnamefont {T.~F.}\ \bibnamefont {Heinz}}, \bibinfo {author}
  {\bibfnamefont {T.}~\bibnamefont {Korn}}, \bibinfo {author} {\bibfnamefont
  {A.}~\bibnamefont {Chernikov}}, \bibinfo {author} {\bibfnamefont
  {E.}~\bibnamefont {Malic}},\ and\ \bibinfo {author} {\bibfnamefont
  {A.}~\bibnamefont {Knorr}},\ }\bibfield  {title} {\bibinfo {title} {Excitonic
  linewidth and coherence lifetime in monolayer transition metal
  dichalcogenides},\ }\href
  {https://doi.org/https://doi.org/10.1038/ncomms13279} {\bibfield  {journal}
  {\bibinfo  {journal} {Nat. Comm.}\ }\textbf {\bibinfo {volume} {7}},\
  \bibinfo {pages} {13279} (\bibinfo {year} {2016})}\BibitemShut {NoStop}%
\bibitem [{\citenamefont {Cadiz}\ \emph {et~al.}(2017)\citenamefont {Cadiz},
  \citenamefont {Courtade}, \citenamefont {Robert}, \citenamefont {Wang},
  \citenamefont {Shen}, \citenamefont {Cai}, \citenamefont {Taniguchi},
  \citenamefont {Watanabe}, \citenamefont {Carrere}, \citenamefont {Lagarde}
  \emph {et~al.}}]{cadiz2017excitonic}%
  \BibitemOpen
  \bibfield  {author} {\bibinfo {author} {\bibfnamefont {F.}~\bibnamefont
  {Cadiz}}, \bibinfo {author} {\bibfnamefont {E.}~\bibnamefont {Courtade}},
  \bibinfo {author} {\bibfnamefont {C.}~\bibnamefont {Robert}}, \bibinfo
  {author} {\bibfnamefont {G.}~\bibnamefont {Wang}}, \bibinfo {author}
  {\bibfnamefont {Y.}~\bibnamefont {Shen}}, \bibinfo {author} {\bibfnamefont
  {H.}~\bibnamefont {Cai}}, \bibinfo {author} {\bibfnamefont {T.}~\bibnamefont
  {Taniguchi}}, \bibinfo {author} {\bibfnamefont {K.}~\bibnamefont {Watanabe}},
  \bibinfo {author} {\bibfnamefont {H.}~\bibnamefont {Carrere}}, \bibinfo
  {author} {\bibfnamefont {D.}~\bibnamefont {Lagarde}}, \emph {et~al.},\
  }\bibfield  {title} {\bibinfo {title} {Excitonic linewidth approaching the
  homogeneous limit in mos 2-based van der waals heterostructures},\ }\href
  {https://doi.org/10.1103/PhysRevX.7.021026} {\bibfield  {journal} {\bibinfo
  {journal} {Physical Review X}\ }\textbf {\bibinfo {volume} {7}},\ \bibinfo
  {pages} {021026} (\bibinfo {year} {2017})}\BibitemShut {NoStop}%
\bibitem [{\citenamefont {Chen}\ \emph {et~al.}(2018)\citenamefont {Chen},
  \citenamefont {Palummo}, \citenamefont {Sangalli},\ and\ \citenamefont
  {Bernardi}}]{Chen2018}%
  \BibitemOpen
  \bibfield  {author} {\bibinfo {author} {\bibfnamefont {H.-Y.}\ \bibnamefont
  {Chen}}, \bibinfo {author} {\bibfnamefont {M.}~\bibnamefont {Palummo}},
  \bibinfo {author} {\bibfnamefont {D.}~\bibnamefont {Sangalli}},\ and\
  \bibinfo {author} {\bibfnamefont {M.}~\bibnamefont {Bernardi}},\ }\bibfield
  {title} {\bibinfo {title} {Theory and ab initio computation of the
  anisotropic light emission in monolayer transition metal dichalcogenides},\
  }\href {https://doi.org/10.1021/acs.nanolett.8b01114} {\bibfield  {journal}
  {\bibinfo  {journal} {Nano Letters}\ }\textbf {\bibinfo {volume} {18}},\
  \bibinfo {pages} {3839} (\bibinfo {year} {2018})}\BibitemShut {NoStop}%
\bibitem [{\citenamefont {Chen}\ \emph {et~al.}(2019)\citenamefont {Chen},
  \citenamefont {Jhalani}, \citenamefont {Palummo},\ and\ \citenamefont
  {Bernardi}}]{Chen2019}%
  \BibitemOpen
  \bibfield  {author} {\bibinfo {author} {\bibfnamefont {H.-Y.}\ \bibnamefont
  {Chen}}, \bibinfo {author} {\bibfnamefont {V.~A.}\ \bibnamefont {Jhalani}},
  \bibinfo {author} {\bibfnamefont {M.}~\bibnamefont {Palummo}},\ and\ \bibinfo
  {author} {\bibfnamefont {M.}~\bibnamefont {Bernardi}},\ }\bibfield  {title}
  {\bibinfo {title} {Ab initio calculations of exciton radiative lifetimes in
  bulk crystals, nanostructures, and molecules},\ }\href
  {https://doi.org/10.1103/PhysRevB.100.075135} {\bibfield  {journal} {\bibinfo
   {journal} {Phys. Rev. B}\ }\textbf {\bibinfo {volume} {100}},\ \bibinfo
  {pages} {075135} (\bibinfo {year} {2019})}\BibitemShut {NoStop}%
\bibitem [{\citenamefont {Palummo}\ \emph {et~al.}(2015)\citenamefont
  {Palummo}, \citenamefont {Bernardi},\ and\ \citenamefont
  {Grossman}}]{Palummo2015}%
  \BibitemOpen
  \bibfield  {author} {\bibinfo {author} {\bibfnamefont {M.}~\bibnamefont
  {Palummo}}, \bibinfo {author} {\bibfnamefont {M.}~\bibnamefont {Bernardi}},\
  and\ \bibinfo {author} {\bibfnamefont {J.~C.}\ \bibnamefont {Grossman}},\
  }\bibfield  {title} {\bibinfo {title} {Exciton radiative lifetimes in
  two-dimensional transition metal dichalcogenides},\ }\href
  {https://doi.org/https://doi.org/10.1021/nl503799t} {\bibfield  {journal}
  {\bibinfo  {journal} {Nano Lett.}\ }\textbf {\bibinfo {volume} {5}},\
  \bibinfo {pages} {2794} (\bibinfo {year} {2015})}\BibitemShut {NoStop}%
\bibitem [{\citenamefont {Kumar}\ \emph {et~al.}(2024)\citenamefont {Kumar},
  \citenamefont {Kolos}, \citenamefont {Bhattacharya},\ and\ \citenamefont
  {Karlický}}]{Nilesh2024}%
  \BibitemOpen
  \bibfield  {author} {\bibinfo {author} {\bibfnamefont {N.}~\bibnamefont
  {Kumar}}, \bibinfo {author} {\bibfnamefont {M.}~\bibnamefont {Kolos}},
  \bibinfo {author} {\bibfnamefont {S.}~\bibnamefont {Bhattacharya}},\ and\
  \bibinfo {author} {\bibfnamefont {F.}~\bibnamefont {Karlický}},\ }\bibfield
  {title} {\bibinfo {title} {{Excitons, optical spectra, and electronic
  properties of semiconducting Hf-based MXenes}},\ }\href
  {https://doi.org/10.1063/5.0197238} {\bibfield  {journal} {\bibinfo
  {journal} {The Journal of Chemical Physics}\ }\textbf {\bibinfo {volume}
  {160}},\ \bibinfo {pages} {124707} (\bibinfo {year} {2024})}\BibitemShut
  {NoStop}%
\bibitem [{\citenamefont {Bange}\ \emph {et~al.}(2023)\citenamefont {Bange},
  \citenamefont {Werner}, \citenamefont {Schmitt}, \citenamefont {Bennecke},
  \citenamefont {Meneghini}, \citenamefont {AlMutairi}, \citenamefont
  {Merboldt}, \citenamefont {Watanabe}, \citenamefont {Taniguchi},
  \citenamefont {Steil}, \citenamefont {Steil}, \citenamefont {Weitz},
  \citenamefont {Hofmann}, \citenamefont {Jansen}, \citenamefont {Brem},
  \citenamefont {Malic}, \citenamefont {Reutzel},\ and\ \citenamefont
  {Mathias}}]{Bange_2023}%
  \BibitemOpen
  \bibfield  {author} {\bibinfo {author} {\bibfnamefont {J.~P.}\ \bibnamefont
  {Bange}}, \bibinfo {author} {\bibfnamefont {P.}~\bibnamefont {Werner}},
  \bibinfo {author} {\bibfnamefont {D.}~\bibnamefont {Schmitt}}, \bibinfo
  {author} {\bibfnamefont {W.}~\bibnamefont {Bennecke}}, \bibinfo {author}
  {\bibfnamefont {G.}~\bibnamefont {Meneghini}}, \bibinfo {author}
  {\bibfnamefont {A.}~\bibnamefont {AlMutairi}}, \bibinfo {author}
  {\bibfnamefont {M.}~\bibnamefont {Merboldt}}, \bibinfo {author}
  {\bibfnamefont {K.}~\bibnamefont {Watanabe}}, \bibinfo {author}
  {\bibfnamefont {T.}~\bibnamefont {Taniguchi}}, \bibinfo {author}
  {\bibfnamefont {S.}~\bibnamefont {Steil}}, \bibinfo {author} {\bibfnamefont
  {D.}~\bibnamefont {Steil}}, \bibinfo {author} {\bibfnamefont {R.~T.}\
  \bibnamefont {Weitz}}, \bibinfo {author} {\bibfnamefont {S.}~\bibnamefont
  {Hofmann}}, \bibinfo {author} {\bibfnamefont {G.~S.~M.}\ \bibnamefont
  {Jansen}}, \bibinfo {author} {\bibfnamefont {S.}~\bibnamefont {Brem}},
  \bibinfo {author} {\bibfnamefont {E.}~\bibnamefont {Malic}}, \bibinfo
  {author} {\bibfnamefont {M.}~\bibnamefont {Reutzel}},\ and\ \bibinfo {author}
  {\bibfnamefont {S.}~\bibnamefont {Mathias}},\ }\bibfield  {title} {\bibinfo
  {title} {Ultrafast dynamics of bright and dark excitons in monolayer wse2 and
  heterobilayer wse2/mos2},\ }\href {https://doi.org/10.1088/2053-1583/ace067}
  {\bibfield  {journal} {\bibinfo  {journal} {2D Materials}\ }\textbf {\bibinfo
  {volume} {10}},\ \bibinfo {pages} {035039} (\bibinfo {year}
  {2023})}\BibitemShut {NoStop}%
\bibitem [{\citenamefont {PALEARI}(2019)}]{Paleari2019}%
  \BibitemOpen
  \bibfield  {author} {\bibinfo {author} {\bibfnamefont {F.}~\bibnamefont
  {PALEARI}},\ }\emph {\bibinfo {title} {First-principles approaches to the
  description of indirect absorption and luminescence spectroscopy:
  exciton-phonon coupling in hexagonal boron nitride}},\ \href
  {https://doi.org/https://orbilu.uni.lu/handle/10993/41058} {Ph.D. thesis},\
  \bibinfo  {school} {Unilu - University of Luxembourg, Luxembourg} (\bibinfo
  {year} {2019})\BibitemShut {NoStop}%
\bibitem [{\citenamefont {Lechifflart}\ \emph {et~al.}(2019)\citenamefont
  {Lechifflart}, \citenamefont {Paleari}, \citenamefont {Sangalli},\ and\
  \citenamefont {Attaccalite}}]{Lechifflart2023}%
  \BibitemOpen
  \bibfield  {author} {\bibinfo {author} {\bibfnamefont {P.}~\bibnamefont
  {Lechifflart}}, \bibinfo {author} {\bibfnamefont {F.}~\bibnamefont
  {Paleari}}, \bibinfo {author} {\bibfnamefont {D.}~\bibnamefont {Sangalli}},\
  and\ \bibinfo {author} {\bibfnamefont {C.}~\bibnamefont {Attaccalite}},\
  }\bibfield  {title} {\bibinfo {title} {First-principles study of luminescence
  in hexagonal boron nitride single layer: Exciton-phonon coupling and the role
  of substrate},\ }\href
  {https://doi.org/https://doi.org/10.1103/PhysRevMaterials.7.024006}
  {\bibfield  {journal} {\bibinfo  {journal} {Phys. Rev. Materials}\ }\textbf
  {\bibinfo {volume} {7}},\ \bibinfo {pages} {024006} (\bibinfo {year}
  {2019})}\BibitemShut {NoStop}%
\end{thebibliography}%

\end{document}